\documentclass[10pt,journal]{IEEEtran}

\usepackage{graphicx}
\DeclareGraphicsExtensions{.png,.eps,.ps,.pdf,.jpg}
\usepackage{url}
\usepackage{color}
\usepackage{cite}
\usepackage{acronym}
\usepackage{epstopdf}
\usepackage{bm}
\usepackage{booktabs}
\usepackage{amsmath}
\usepackage{amssymb}
\usepackage{siunitx}
\usepackage{cancel}
\usepackage{caption}
\usepackage{multirow}
\usepackage{mathrsfs}
\usepackage{subcaption}
\usepackage[normalem]{ulem}
\usepackage[utf8]{inputenc}

\usepackage{amsmath}
\usepackage[table,xcdraw]{xcolor}
\usepackage{pgfplots}
\usepackage{graphicx}
\usepackage{tabularx}
\usepackage{colortbl}
\usepackage{xfrac}
\newcommand{\ra}[1]{\renewcommand{\arraystretch}{#1}}
\usepackage[]{algorithm2e,setspace}
\usepackage{enumitem}
\usepackage{fixltx2e}

\newcommand\tabitem{\makebox[1em][r]{\textbullet~}}

\newcommand{\subparagraph}{}
\usepackage{titlesec}
\titlespacing\section{0pt}{5pt}{1pt}[0pt]
\titlespacing\subsection{0pt}{5pt}{1pt}[0pt]
\titlespacing\subsubsection{0pt}{5pt}{1pt}[0pt]

\acrodef{IP}[IP]{Internet Protocol}
\acrodef{TCP}[TCP]{Transmission Control Protocol}
\acrodef{UDP}[UDP]{User Datagram Protocol}
\acrodef{CDMA}[CDMA]{Code Division Multiple Access}
\acrodef{ML}[ML]{Machine Learning}
\acrodef{SVM}[SVM]{Support Vector Machine}
\acrodef{CDR}[CDR]{Call Data Record}
\acrodef{app}[app]{application}
\acrodef{LTE}[LTE]{Long Term Evolution}
\acrodef{PDCCH}[PDCCH]{Physical Downlink Control CHannel}
\acrodef{eNodeB}[eNodeB]{Base Station}
\acrodef{DCI}[DCI]{Downlink Control Information}
\acrodef{TTI}[TTI]{Transmission Time Interval}
\acrodef{MCS}[MCS]{Modulation and Coding Scheme}
\acrodef{CNN}[CNN]{Convolutional Neural Network}
\acrodef{SDR}[SDR]{Software-Defined Radio}
\acrodef{UE}[UE]{User Equipment}
\acrodef{TBS}[TBS]{Transport Block Size}
\acrodef{RNTI}[RNTI]{Radio Network Temporary Identifier}
\acrodef{LSTM}[LSTM]{Long Short-Term Memory}
\acrodef{MLP}[MLP]{Multilayer Perceptron}
\acrodef{ReLU}[ReLU]{Rectified Linear Unit}
\acrodef{RNN}[RNN]{Recurrent Neural Network}
\acrodef{NN}[NN]{Neural Network}
\acrodef{MC}[MC]{Memory Cell}
\acrodef{GP}[GP]{Gaussian Process}
\acrodef{RB}[RB]{Resource Block}
\acrodef{RAN}[RAN]{Radio Access Network}
\acrodef{HTTP}[HTTP]{Hypertext Transfer Protocol}
\acrodef{GAN}[GAN]{Generative-Adversarial Networks}
\acrodef{SSD}[SSD]{Solid State Storage}
\acrodef{HMM}[HMM]{Hidden Markov Model}
\acrodef{RF}[RF]{Random Forests}

\graphicspath{{./figures/}}

\begin{document}

\title{Mobile Traffic Classification through Physical Channel Fingerprinting: a Deep Learning Approach}  

\author{\IEEEauthorblockN{Hoang Duy Trinh\IEEEauthorrefmark{1}, Angel Fernandez Gambin\IEEEauthorrefmark{2}, Lorenza Giupponi\IEEEauthorrefmark{1}, Michele Rossi\IEEEauthorrefmark{2} and Paolo Dini\IEEEauthorrefmark{1}}
	\thanks{\IEEEauthorrefmark{1}CTTC/CERCA, Av. Carl Friedrich Gauss, 7, 08860, Castelldefels, Barcelona, Spain \{hoangduy.trinh, lorenza.giupponi, paolo.dini\}@cttc.es, \IEEEauthorrefmark{2}DEI, University of Padova, Via G. Gradenigo, 6/B, 35131 Padova, Italy. \{afgambin, rossi\}@dei.unipd.it. 
	
This work has received funding from the European Union Horizon 2020 research and innovation programme under the Marie Sklodowska-Curie grant agreement No. 675891 (SCAVENGE), by the Spanish Government under project TEC2017-88373-R (5G-REFINE) and has been supported, in part, by MIUR (Italian Ministry of Education, University and Research) through the initiative ``Departments of Excellence'' (Law 232/2016). } \vspace{-0.7cm}}



\maketitle
\vspace{-1pt}

\begin{abstract} 
The automatic classification of applications and services is an invaluable feature for new generation mobile networks. Here, we propose and validate algorithms to perform this task, at {\it runtime}, from the {\it raw physical channel} of an {\it operative mobile network}, without having to decode and/or decrypt the transmitted flows. Towards this, we decode Downlink Control Information (DCI) messages carried within the LTE Physical Downlink Control CHannel (PDCCH). DCI messages are sent by the radio cell in clear text and, in this paper, are utilized to classify the applications and services executed at the connected mobile terminals. Two datasets are collected through a large measurement campaign: one labeled, used to train the classification algorithms, and one unlabeled, collected from four radio cells in the metropolitan area of Barcelona, in Spain. Among other approaches, our Convolutional Neural Network (CNN) classifier provides the highest classification accuracy of $99\%$. The CNN classifier is then augmented with the capability of rejecting sessions whose patterns do not conform to those learned during the training phase, and is subsequently utilized to attain a fine grained decomposition of the traffic for the four monitored radio cells, in an {\it online} and {\it unsupervised} fashion. 
\end{abstract}

\begin{IEEEkeywords}
Traffic Classification, Traffic Modeling, Mobile Networks, LTE, 5G, Machine Learning, Neural Networks, Deep Learning, Data Analytics. 
\end{IEEEkeywords}


\section{Introduction}
\label{sec:intro}

\IEEEPARstart{W}{ireless} mobile technology is advancing at a fast pace, through better monitor resolutions, larger memories, higher communication speeds, etc., and, with that, more requirements in terms of supported data rates~\cite{ericsson,ciscovni}, new services, and a higher network responsiveness across diverse physical contexts~\cite{chen2014requirements}.

In this work, we design and evaluate, via a \mbox{proof-of-concept} implementation, \mbox{non-intrusive} tools for the online estimation of \ac{LTE} cellular activity, i.e., the type of traffic that users exchange with their serving base station. As we quantify, our technology allows one to infer {\it with high accuracy} the service (e.g., audio, video, etc.) and the application (e.g., Skype, Vimeo, You Tube, etc.) that are being used by the connected mobile users. This is accomplished by decoding LTE downlink/uplink control channel messages (i.e., \mbox{radio-link} level data), which are transmitted in clear text and without breaking any security protocol (encryption). We believe that this brings a high value along several dimensions:
\begin{itemize}
\item first, our tool permits a better understanding of spectrum needs across time and space. Note that there are limited means to \mbox{non-intrusively} monitor user density and traffic demand in \mbox{real-time}. These measurements are key for the correct dimensioning of future mobile systems, and the investigation of new data communication and processing techniques at the network edge. In fact, although network operators gather such data through their base stations, they seldom release it due to privacy concerns, so these measures are almost never used for research purposes by university, public or private research labs. We only found a limited number of datasets available including mobile data traffic~\cite{Laurila2012}\cite{Barlacchi2015}, but they are often incomplete, poorly documented and contain aggregated data, often averaged hourly or across a certain geographic area. Our technique allows the extraction of data flows with a granularity of one second, isolating the amount of data, the type of service and application exploited by each mobile user within an LTE cell. We believe that this has a high value for researchers within the mobile networking space, who will be able to create their own {\it rich} datasets for research purposes, to spur the advancement of future edge communication (and computing) technology; 
\item second, our work sheds new light on possible attacks that may be carried out by exploiting the LTE control channel data. In fact, being able to track user density across time and space by just deploying a number of sniffers within a city may be also exploited maliciously, for example to infer details about mobility and user density. While such analysis, so far, has been carried out with data from network operators, e.g., the Telecom Italia mobile challenge~\cite{Reades2009}, our tool allows the extraction of fine grained LTE data in full autonomy. This may represent a privacy concern.  
\end{itemize}

A large body of work exists in the area of mobile traffic classification (see Section~\ref{sec:related} for an in depth discussion of the related work). The key challenge of existing classification algorithms consists in the identification, and in the subsequent computation, of a number of {\it representative features}. These features are then used to train algorithms that classify the data flows at runtime. Most of the surveyed approaches leverage some {\it domain knowledge}, which is utilized to {\it manually} obtain the feature set, i.e., by a skilled human expert. However, the use of deep learning techniques has recently paved the way to automatic feature discovery and extraction, often leading to superior performance. For example, in~\cite{aceto2018mobile} encrypted traffic is categorized through deep learning architectures, proving their superior performance with respect to shallow neural network classifiers. The authors of~\cite{zhang2017zipnet} present a mobile traffic \mbox{super-resolution} technique to infer narrowly localized traffic consumption from coarse measurements: a \mbox{deep-learning} architecture combining Zipper Network (ZipNet) and Generative Adversarial neural Network (GAN) models is proposed to accurately reconstruct \mbox{spatio-temporal} traffic dynamics from measurements taken at low resolution. In~\cite{chen2016automatic}, the identification of mobile apps is carried out by automatically extracting features from labeled packets through \acp{CNN}, which are trained using raw \ac{HTTP} requests, achieving a high classification accuracy. We stress that the work in these papers, as the majority of the other techniques discussed in Section~\ref{sec:related}, use statistical features obtained from application or \mbox{\ac{IP}} level information for both service and app identification, along with UDP/TCP port numbers.

The solution here presented sharply departs from previous approaches, as it performs highly accurate traffic classification directly from {\it radio-link} level data, without requiring any prior knowledge and without having to decode and/or decrypt the transmitted data flows. The proposed classifiers enable fully automated, over the air, and detailed traffic profiling in mobile cellular systems, making it possible to infer, in an online fashion, the radio resource usage of typical service classes. To do this, we leverage OWL~\cite{bui2016owl}, a tool that allows decoding the \ac{LTE} \ac{PDCCH}, where control information is exchanged between the \ac{LTE} \ac{eNodeB} and the connected \acp{UE}. Specifically, we decode the \ac{DCI} messages carried in the \ac{PDCCH}, which contain \mbox{radio-link} level settings for the user communication (e.g., modulation and coding scheme, transport block size, allocated resource blocks). From \ac{DCI} data, we create two datasets
\begin{itemize}
\item[\textbf{1)}] a \emph{labeled dataset}, used to {\it train} service and app classification algorithms, where labeling is made possible by injecting an easily identifiable watermark into the application flows generated by a  terminal under our control; 
\item[\textbf{2)}] an \emph{unlabeled dataset}, used for {\it traffic profiling} purposes, which is populated by monitoring, for a full month, mobile traffic from four operative radio cell sites with different demographic characteristics within the metropolitan area of Barcelona, in Spain.
\end{itemize}

For the traffic analysis, we focus on a few services and applications that dominate the radio resource usage, but the approach can be readily extended to further services and applications. We directly use raw DCI data as input into deep learning classifiers (automatic feature extraction), achieving accuracies as high as $98$\% for both mobile service and app identification tasks. Moreover, we propose a novel technique to use our best classifier in {\it unsupervised settings}, to profile the mobile traffic from operative radio cell sites at runtime. Our tool allows for a {\it fine grained} and {\it automated} analysis of user traffic from real deployments, using radio link messages transmitted over the \ac{PDCCH}. The developed classification algorithms, as well as our experimental results are highly novel within the traffic monitoring literature, which only provides hourly and {\it aggregated} measures for typical days~\cite{earth-D23}\cite{Xu2017understanding}, and where traffic profiling is performed from UDP/TCP, IP or above IP flows, e.g.,~\cite{aceto2018mobile, zhang2017zipnet, chen2016automatic}.

In summary, the original contributions of this work are

\begin{itemize}
	\item \textit{Mobile Data Labeling:} we present an original and effective approach to automatically label \ac{LTE} PDCCH \ac{DCI} data traces. This approach is utilized for six mobile apps, to create a unique correspondence between the software programs (the apps) and the session identifiers that were assigned to them by the eNodeB. The result is a {\it labeled dataset} of real \ac{DCI} data from selected applications, i.e., YouTube, Vimeo, Spotify, Google Music, Skype and WhatsApp video calls.
	\item \textit{Classification and Benchmarks:} we tailor deep artificial Neural Networks (NNs), namely \mbox{Multi-Layer} Perceptron (MLP), \acp{RNN} and \acp{CNN}, to perform classification tasks for mobile services and app identification on the labeled dataset. Moreover, we compare their performance against a number of (benchmark) \mbox{state-of-the-art} classifiers. 
	\item \textit{Mobile Data Collection from an Operative Mobile Network:} we collect real \ac{LTE} PDCCH \ac{DCI} data traces, with a time granularity of $1$~ms, from four \acp{eNodeB} located in the metropolitan area of Barcelona, in Spain. Each of these datasets has a duration of $1$~month.
	\item \textit{Mobile Service Profiling from Unlabeled Data:} the \ac{CNN} classifier, which is found to be the best among all \ac{NN} schemes, is augmented with the capability of rejecting {\it out of distribution} sessions, i.e., sessions whose statistical behavior departs from those learned during the training phase. This makes it possible to use it with unlabeled traffic (online and unsupervised settings). The augmented \ac{CNN} classifier rejects those sessions for which it is uncertain, providing a robust classification outcome. Through its use, the four selected \acp{eNodeB} are monitored, getting a fine grained traffic decomposition. 
\end{itemize}

The paper is organized as follows. Section~\ref{sec:dataset} presents the experimental framework and the proposed methodology to obtain the two datasets. Section~\ref{sec:sl} introduces the two classification problems, namely, service and app identification and presents the classification algorithms. The performance of such classification algorithms is assessed in Section~\ref{sec:results}. In Section~\ref{sec:traffic_profilig}, the \ac{CNN} classifier is augmented with the capability of rejecting out of distribution sessions. Thus, the mobile traffic from four selected cell sites of an operative mobile network in Spain is decomposed over a full day. The related work on mobile traffic classification is reviewed in Section~\ref{sec:related}, and some concluding remarks are provided in Section~\ref{sec:conclusions}.

\begin{figure*}[t!]
\centering
\includegraphics[width=0.8\textwidth]{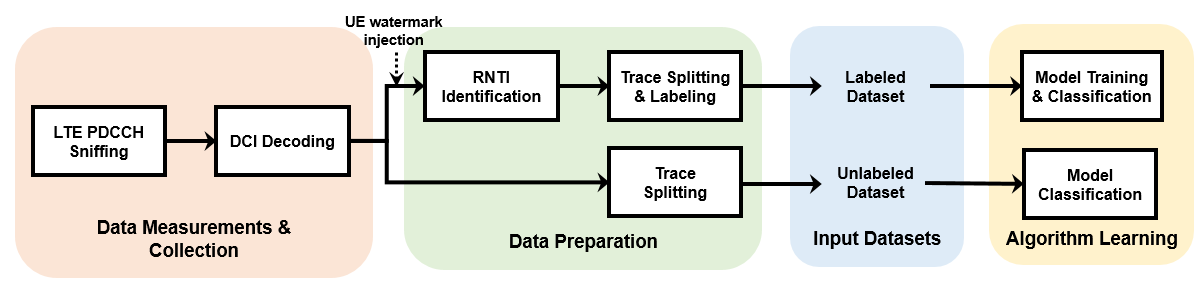}
\caption{Experimental framework adopted for the creation of the unlabeled and labeled datasets. }
\label{fig:framework_diagram}
\end{figure*}

\section{Dataset Creation}
\label{sec:dataset}

Fig.~\ref{fig:framework_diagram} shows the different building blocks of the experimental framework that has been developed to populate the unlabeled and labeled datasets. Briefly, the {\it data measurement and collection} block acquires data from the LTE \ac{PDCCH} channel to extract the relevant \ac{DCI} information. {\it Data preparation}, instead, processes the gathered \ac{DCI} data so that it can be used for training and classification purposes.

\subsection{Data Collection System}

In LTE, the eNodeB communicates scheduling information to the connected \acp{UE} through the DCI messages that are carried within the PDCCH with a time granularity of $1$~ms. While the actual user content is sent over {\it encrypted dedicated channels}, i.e., the Physical Uplink/Downlink Shared Channel (PUSCH/PDSCH respectively), the \textbf{PDCCH is transmitted in clear text} and can be decoded. To process DCI data, we have adapted the OWL monitoring tool~\cite{bui2016owl}. A \ac{SDR} has been programmed, acquiring the \ac{PDCCH} via an \mbox{open-source} software sitting on top of the \mbox{srs-LTE} library~\cite{gomez2016srslte}, which makes it possible to synchronize and monitor the channel over a specified \ac{LTE} bandwidth. The SDR is connected to a PC that performs the actual decoding of DCI data: in our experimental settings, we used a low cost Nuand BladeRF x40 \ac{SDR} and an Intel \mbox{mini-NUC}, equipped with an i5 $2.7$~Ghz \mbox{multi-core} processor, $256$~GB \ac{SSD} storage and $18$~GB of RAM. 

\begin{figure*}[t]
	\centering
	\begin{subfigure}[t]{.49\columnwidth}
		\centering
		\includegraphics[width=\columnwidth]{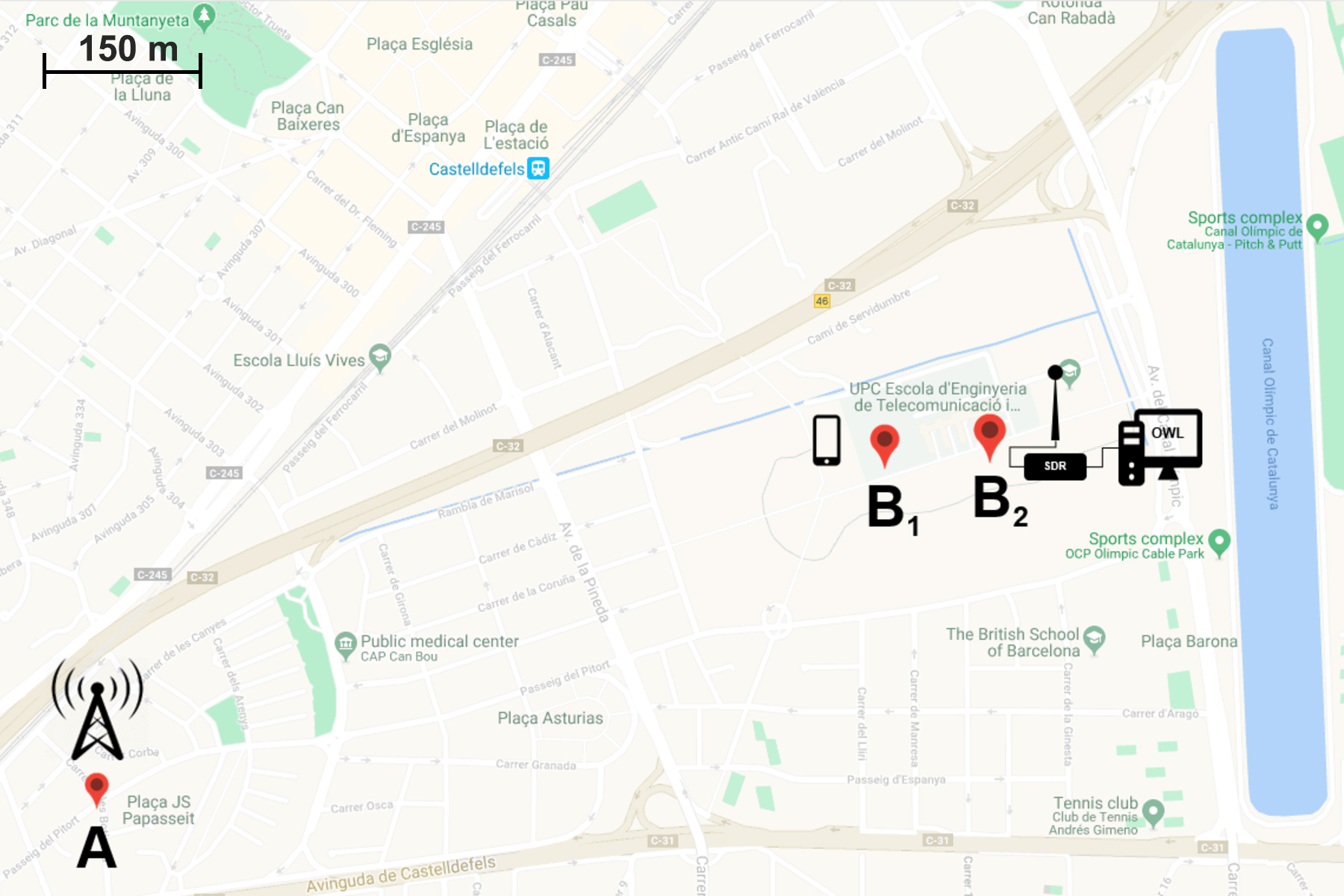}
		\caption{Castelldefels: suburban area with a university campus.\vspace{0.3cm}}
		\label{fig:castefa}
	\end{subfigure}
	\begin{subfigure}[t]{.49\columnwidth}
		\centering
		\includegraphics[width=\columnwidth]{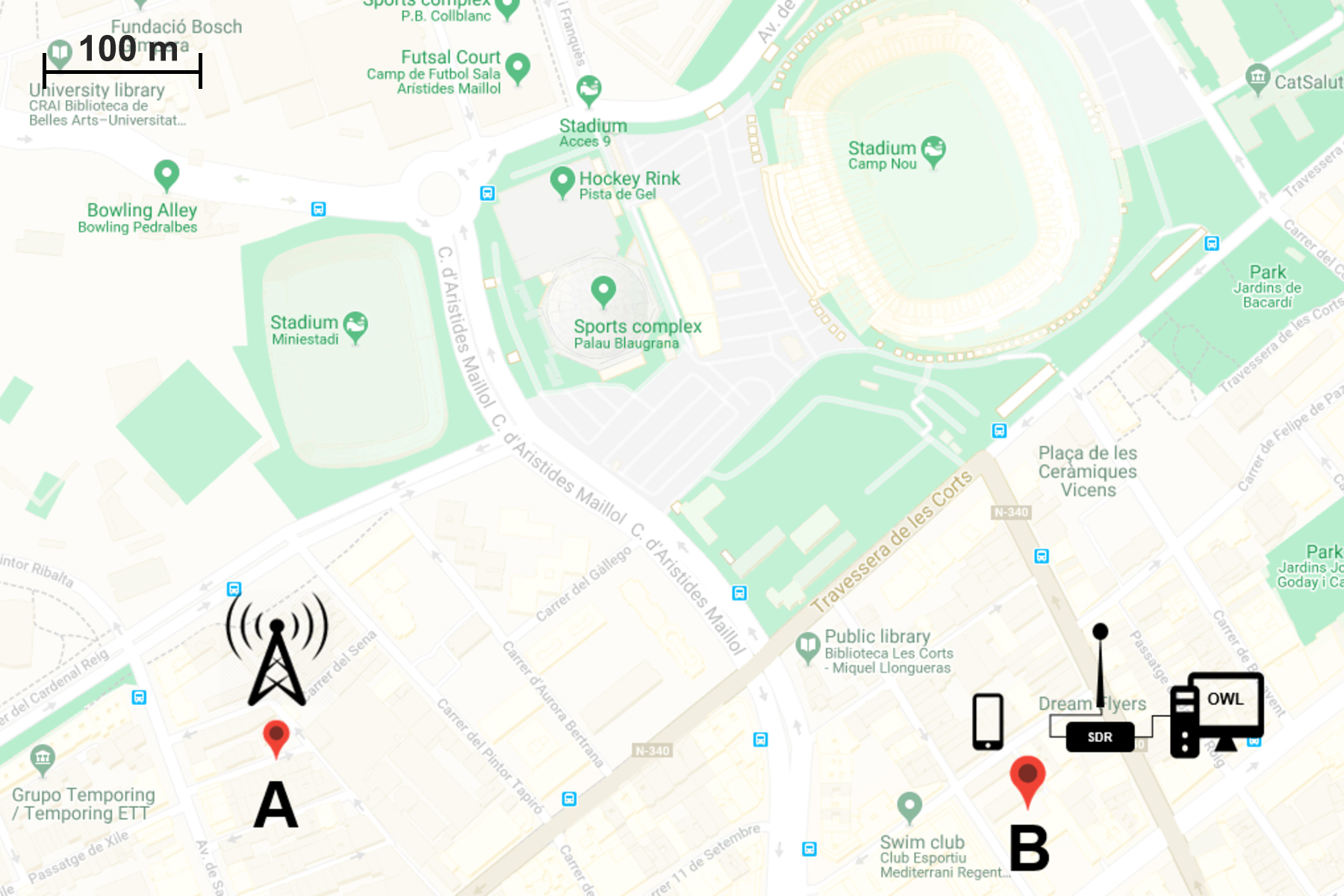}
		\caption{Camp Nou: mainly residential area with Barcelona FC stadium.\vspace{0.3cm}}
		\label{fig:bcn}	
	\end{subfigure}
	\begin{subfigure}[t]{.49\columnwidth}
		\centering
		\includegraphics[width=\columnwidth]{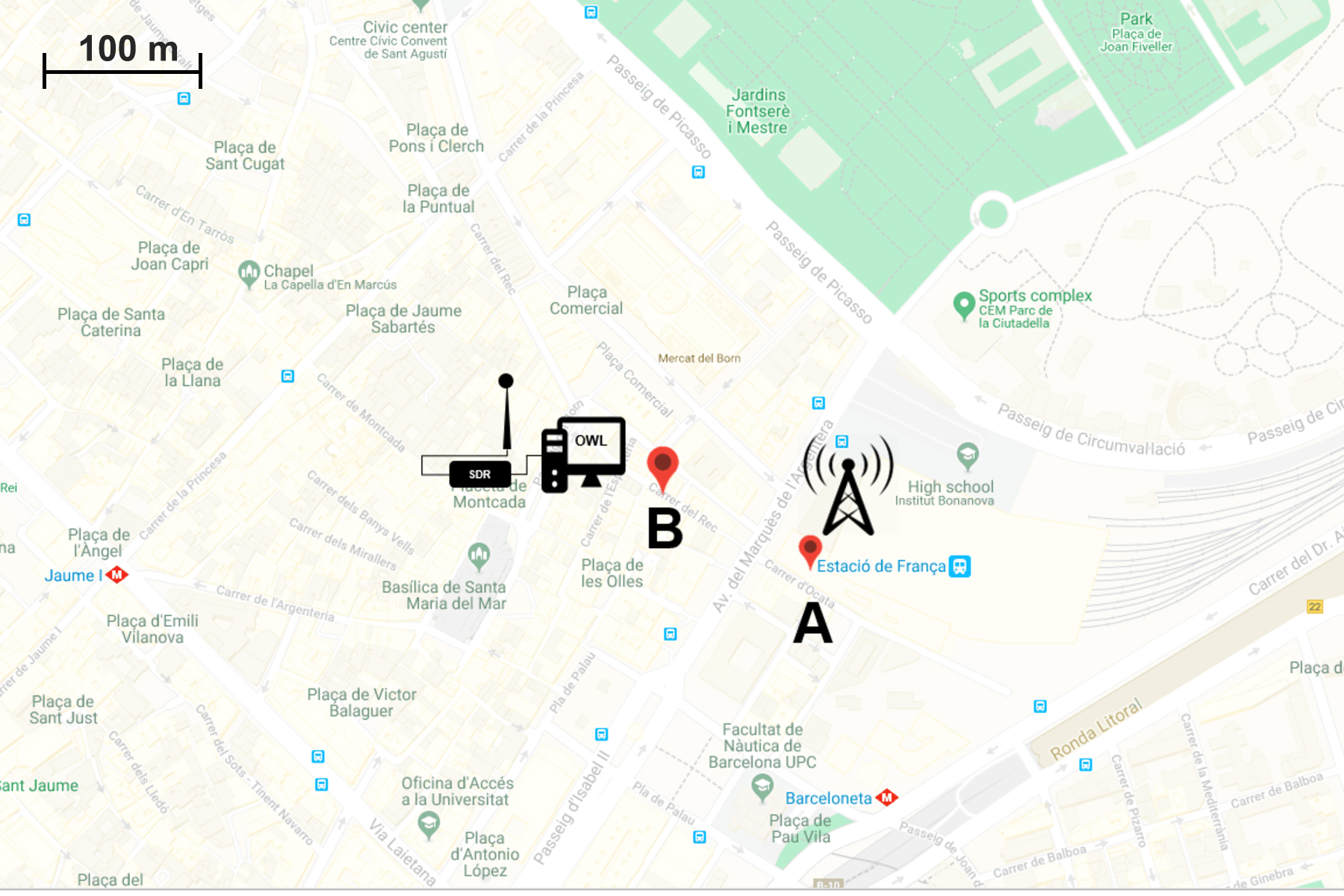}
		\caption{Born: mixed residential, transport and leisure area.}
		\label{fig:born}	
	\end{subfigure}
	\begin{subfigure}[t]{.49\columnwidth}
		\centering
		\includegraphics[width=\columnwidth]{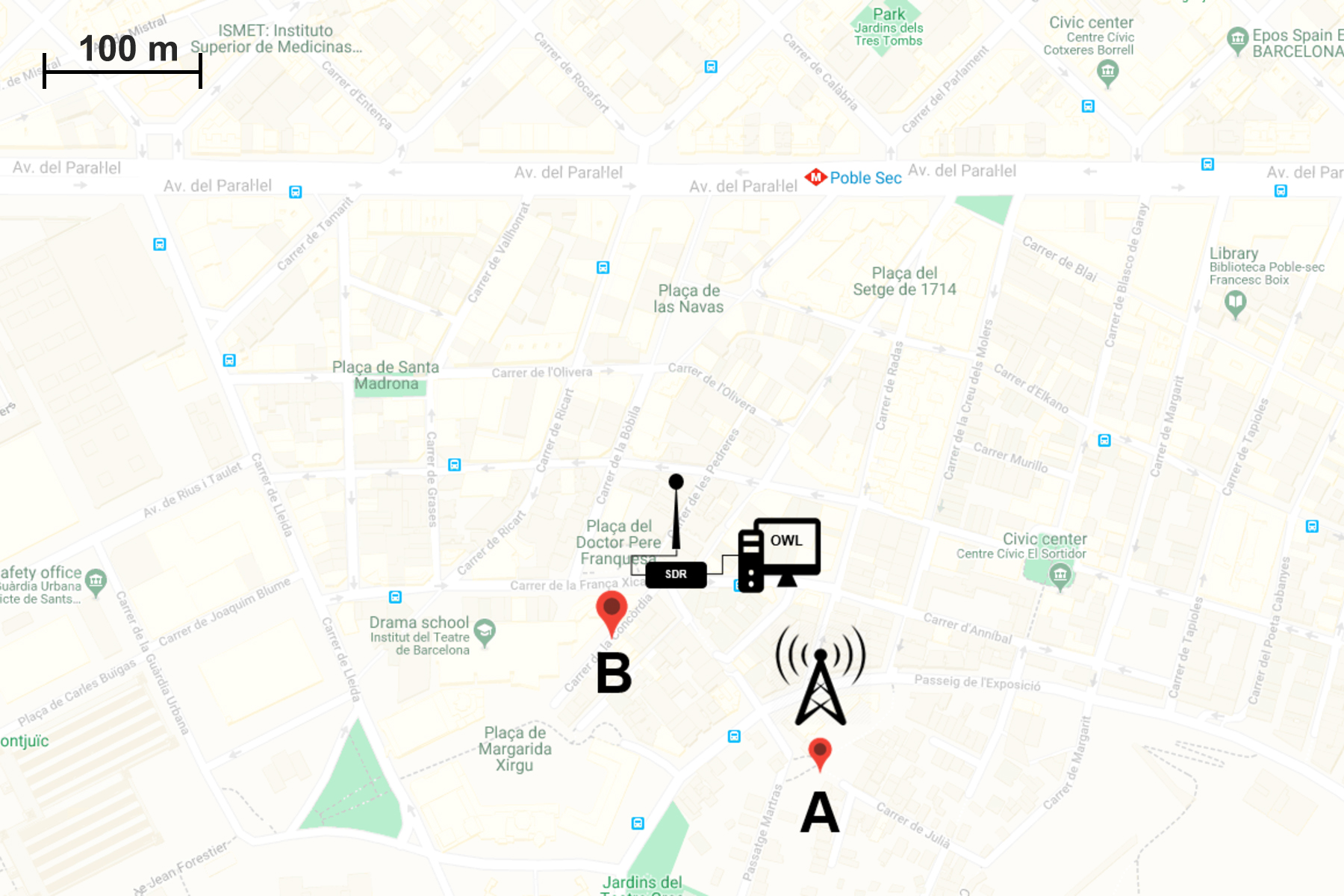}
		\caption{PobleSec: mainly residential area.}
		\label{fig:poblesec}	
	\end{subfigure}
	\centering
	\caption{Maps of Barcelona metropolitan areas where the measurement campaign took place for the creation of unlabeled and labeled datasets. In the maps, the \ac{eNodeB} location is denoted by $A$, whereas the data collection system and the mobile terminal are marked as B. In Castelldefels, the mobile terminal has been placed in two different locations (B$_1$ and B$_2$).}
	\label{fig:maps}
\end{figure*}

Decoded \ac{DCI} messages for a connected \ac{UE} contain the following scheduling information~\cite{etsi}:
\begin{itemize}
	\itemsep0.5em
	\item \ac{RNTI},
	\item \ac{RB} assignment,
	\item \ac{MCS}.
\end{itemize}

\ac{DCI} messages use \acp{RNTI} to specify their destination. \acp{RNTI} are \mbox{$16$-bit} identifiers that are employed to address \acp{UE} in an \ac{LTE} cell. They are used for different purposes such as to broadcast system information \mbox{(SI-RNTI)}, to page a specific UE \mbox{(P-RNTI)}, to carry out a random access procedure \mbox{(RA-RNTI)}, and to identify a connected user, i.e., the cell RNTI \mbox{(C-RNTI)}. In this work, we are interested in the \mbox{C-RNTI}, that is temporarily assigned when the UE is in RRC (Radio Resource Control) CONNECTED state, to uniquely identify it inside the cell. The \mbox{C-RNTI} can take any unreserved value in the range \mbox{[{\tt 0x003D}--{\tt FFF3}].} Once the \mbox{C-RNTI} is assigned to a connected \ac{UE}, the \ac{DCI} information directed to this terminal is sent using this \mbox{C-RNTI}, which is transmitted in clear text as part of the \ac{PDCCH} channel. Hence, knowing the \mbox{C-RNTI} allows tracking a specific connected user within the radio cell. Assuming that the \mbox{C-RNTI} is known (see Section~\ref{sec:labeled_dataset}), the following information about the ongoing communication for this \ac{UE} can be extracted from its \ac{DCI} data:

\begin{itemize}
\item {\it Number of allotted resource blocks:} in \ac{LTE}, a \ac{RB} represents the smallest resource unit that can be allocated to any user. The number of resource blocks that are assigned to a \ac{UE} (N\textsubscript{RB}), is derived based on the \ac{DCI} bitmap.

\item {\it Modulation order and code rate:} the \ac{MCS} is a \mbox{$5$-bit} field that determines the modulation order and the code rate that are used, at the physical layer, for the transmission of data to the UE.

\item {\it \ac{TBS}:} the \ac{TBS} specifies the length of the packet to be sent to the \ac{UE} in the current \ac{TTI}. It is derived by from a lookup table by using \ac{MCS} and N\textsubscript{RB}, see~\cite{etsi}.
\end{itemize}
In this work, we demonstrate that monitoring the downlink and uplink \ac{TBS} information of a given \ac{UE} enables service and app classification with high accuracy.

\subsection{Unlabeled Dataset}
\label{sec:unlabeled_dataset}

Thanks to the just described \ac{DCI} collection system, four cell sites of a Spanish mobile network operator in the metropolitan area of Barcelona have been monitored for a full month. The selected \acp{eNodeB} are located in areas having different demographic characteristics and land uses, so as to diversify the captured traffic in terms of service and app behavior. We have named the datasets according to the corresponding neighborhood: {\it PobleSec} (mainly residential area), {\it Born} (mixed residential, transport and leisure area), {\it Castelldefels} (mixed suburban and campus area), {\it Camp Nou} (mixed residential and stadium area). In total, we have collected more than $68$~GB of \ac{DCI} data from the \ac{LTE} \ac{PDCCH}. Fig.~\ref{fig:maps} shows the locations of the four monitored sites, along with that of the data collection system. After the data collection, the signaling associated with each active \mbox{C-RNTI} is extracted from the \ac{PDCCH} \ac{DCI} data stream, and is prepared for the classifier. During this, we discarded \mbox{short-length} traces, which are mainly due to signaling, paging and background traffic. These accounted for less than $3\%$ of the total traffic in the monitored radio cells.

\subsection{Labeled Dataset}
\label{sec:labeled_dataset}

A {\it labeled} dataset is obtained by running specific services and apps at a mobile terminal under our control, detecting its \mbox{C-RNTI} within the \ac{PDCCH} channel and finally associating the corresponding \ac{DCI} trace with a {\it label}, which links it to the service/app that is executed at the \ac{UE}. Generating data sessions is easy, and boils down to running a specific app from a device that we control, and that is connected to the monitored \ac{eNodeB}. The difficult part is to identify the generated data flow among those carried by the \ac{PDCCH} channel, which contains \ac{DCI} information for {\it all} the connected \acp{UE} within the radio cell. We made this labeling possible by injecting a {\it watermark} into the traffic that we generated by the controlled \ac{UE}, so that it could be easily identified among all other users.

\subsubsection{Data preparation and watermarking} The data preparation procedure is divided into two phases: \textbf{1)} the {\it identification} of the \mbox{C-RNTI} corresponding to the controlled \ac{UE}, \textbf{2)} the {\it extraction} and {\it labeling} of the corresponding \ac{DCI} trace. In the \ac{LTE} \ac{PDCCH} channel, each \ac{UE} is identified by the \mbox{C-RNTI}, which uniquely identifies the mobile terminal within the radio cell. This identifier is {\it temporary}, i.e., it changes after short inactivity periods. This is done to prevent the plain tracking of mobile users, since the \ac{PDCCH} is sent unencrypted. To allow traffic labeling (i.e., user identification), we introduced a {\it watermark} into the traffic generated through our mobile terminal. This watermark amounts to producing, for each application, a regular pattern: any given application (e.g., YouTube) is run for a \mbox{pre-defined} amount of time ($60$~seconds in our measurements), then, a pause interval of fixed duration is inserted before running the app for further $60$~seconds. We loop this over time, obtaining a duty cycled activity pattern that is easily distinguishable from all the other activity traces within the radio cell. Through this watermarking procedure, we could successfully associate our \ac{UE} with the corresponding \mbox{C-RNTI} from the \ac{DCI}. Also, we split the traces into different sessions thanks to the duty cycled pattern, where subsequent sessions are separated by the pause interval (of fixed duration). The {\it label}, corresponding to the application that is being executed at the mobile terminal, was finally associated with the extracted \ac{DCI} data.

In our measurement campaign, we have recorded and labeled about $M=10,000$ mobile sessions, gathering the scheduling information contained in the \ac{DCI} messages for selected apps. We considered three \mbox{data-intensive} services: {\it video streaming}, {\it audio streaming} and {\it \mbox{real-time} video calling}, which represent classes producing a considerable amount of traffic and taking most of the network resources~\cite{ericsson}. For each service type, we chose two popular applications: \textit{Spotify} and \textit{Google Music} for audio, \textit{YouTube} and \textit{Vimeo} for video streaming, while for the video calling we picked two \mbox{instant-messaging} applications, namely, \textit{Skype} and \textit{WhatsApp Messenger}.

A large measurement campaign was conducted to expose the mobile terminal running the selected apps to different {\it radio link conditions}, thus obtaining a comprehensive dataset. In particular, the \ac{UE} was placed into two different locations (termed B$_1$ and B$_2$ in Fig.~\ref{fig:castefa}) within the Castelldefels radio cell to experience different received signal qualities ($-84$~dBm and $-94$~dBm for B$_1$ and B$_2$, respectively), and in the Camp Nou \ac{eNodeB} during football matches, to capture data in high cell load conditions.

\begin{figure*}[t]
	\centering
	\includegraphics[width=2\columnwidth]{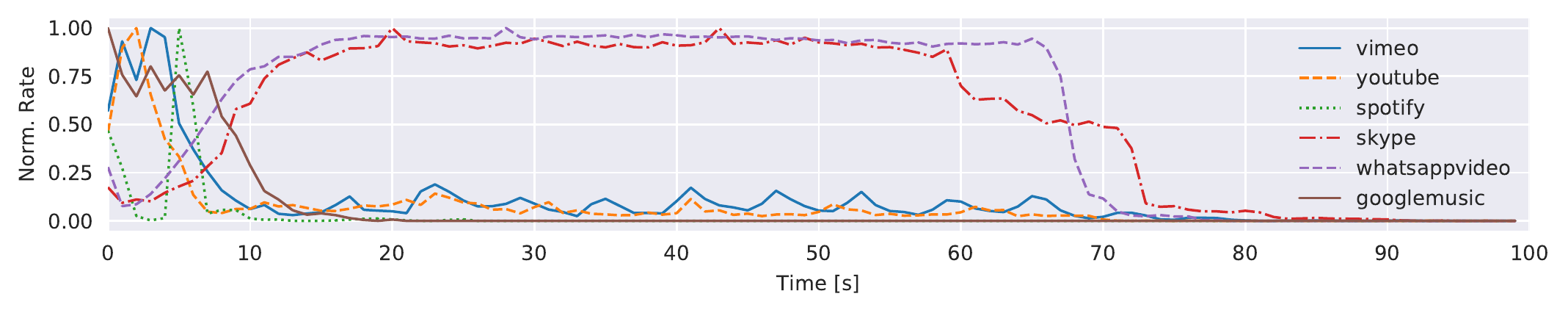}
		\includegraphics[height=4.2cm]{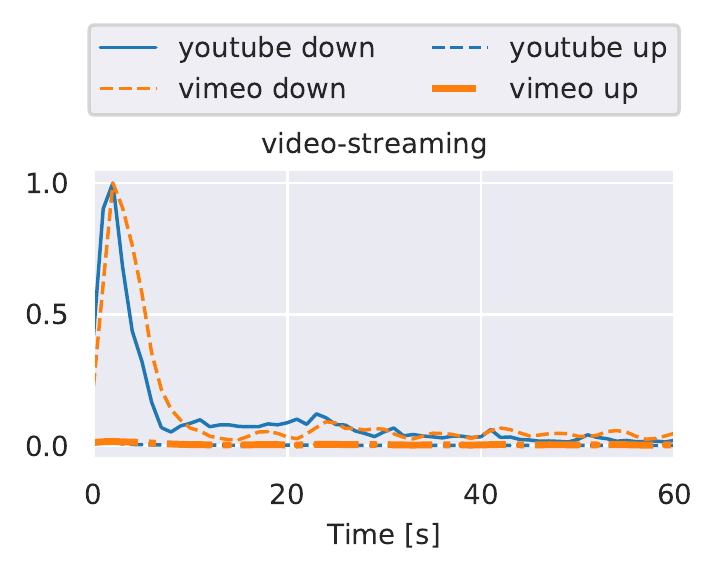}
	\includegraphics[height=4.2cm]{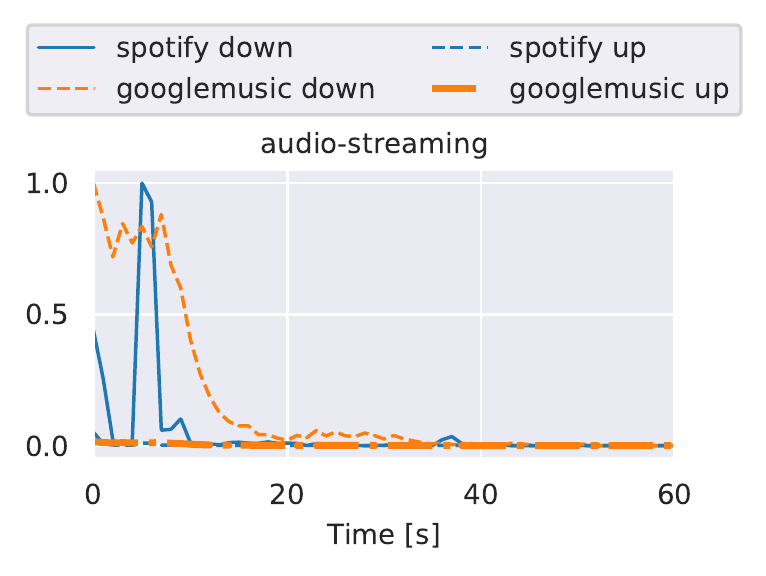}
	\includegraphics[height=4.2cm]{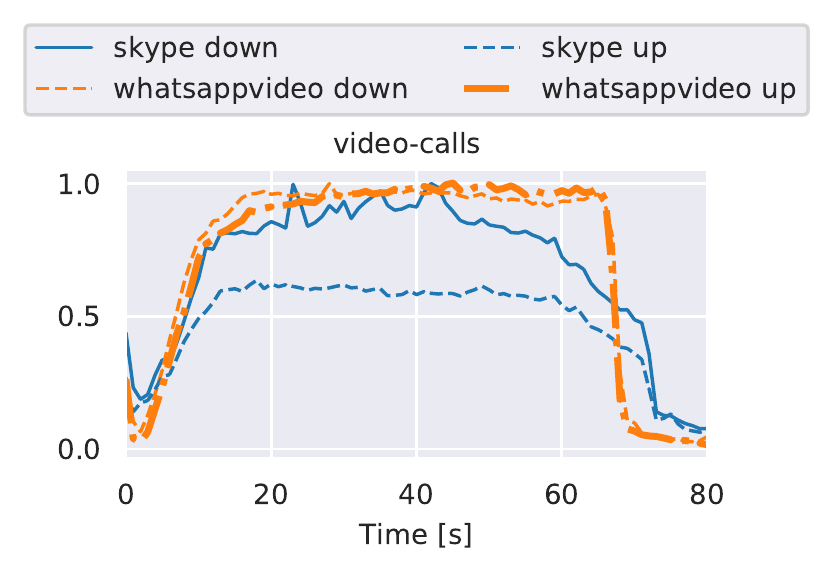}
	\caption{Traffic pattern snapshots showing the normalized data rate for different applications as a function of time.}
	\label{fig:trace_vs_time}
\end{figure*}

Fig.~\ref{fig:trace_vs_time} shows a few radio resource usage patterns collected for the selected apps. Some similarities can be recognized within the same service class. For example, audio and video streaming present similar behaviors. Also, significant differences can be observed between the radio resource usage of real time video calls (\textit{Skype} and \textit{WhatApp Video}) and the other apps. Video and audio streaming applications use up a high amount of radio resources at the beginning of the sessions, buffering most of the content into the terminal memory. Real time video calling, instead, entails a continuous transmission and a more constant usage of radio resources throughout the sessions. Note that the amount of data exchanged in the uplink direction is significant only for this service class, since a video call requires a bidirectional communication.

\subsection{Synchronous and Asynchronous Sessions}

Through the watermarking approach and the splitting procedure, we obtained a labeled dataset, where each session, depending on the service, presents patterns similar to those shown in Fig.~\ref{fig:trace_vs_time}. Assuming that the beginning and the end of each session are known is rather optimistic, as in a real measurement setup we have no means to accurately track these instants. Put it another way, it is unlikely that the LTE PDCCH measurements and the application run on the \ac{UE} will be temporally \textit{synchronized}. Synchronizing the measurement with the beginning of each session would facilitate the classification task, since most of the generated traffic is buffered on the terminal at the beginning, see Fig.~\ref{fig:trace_vs_time}, and this behavior is a distinctive feature that is easy to discriminate. 

\begin{figure}[t!]
	\centering
	\includegraphics[width=.9\columnwidth]
	{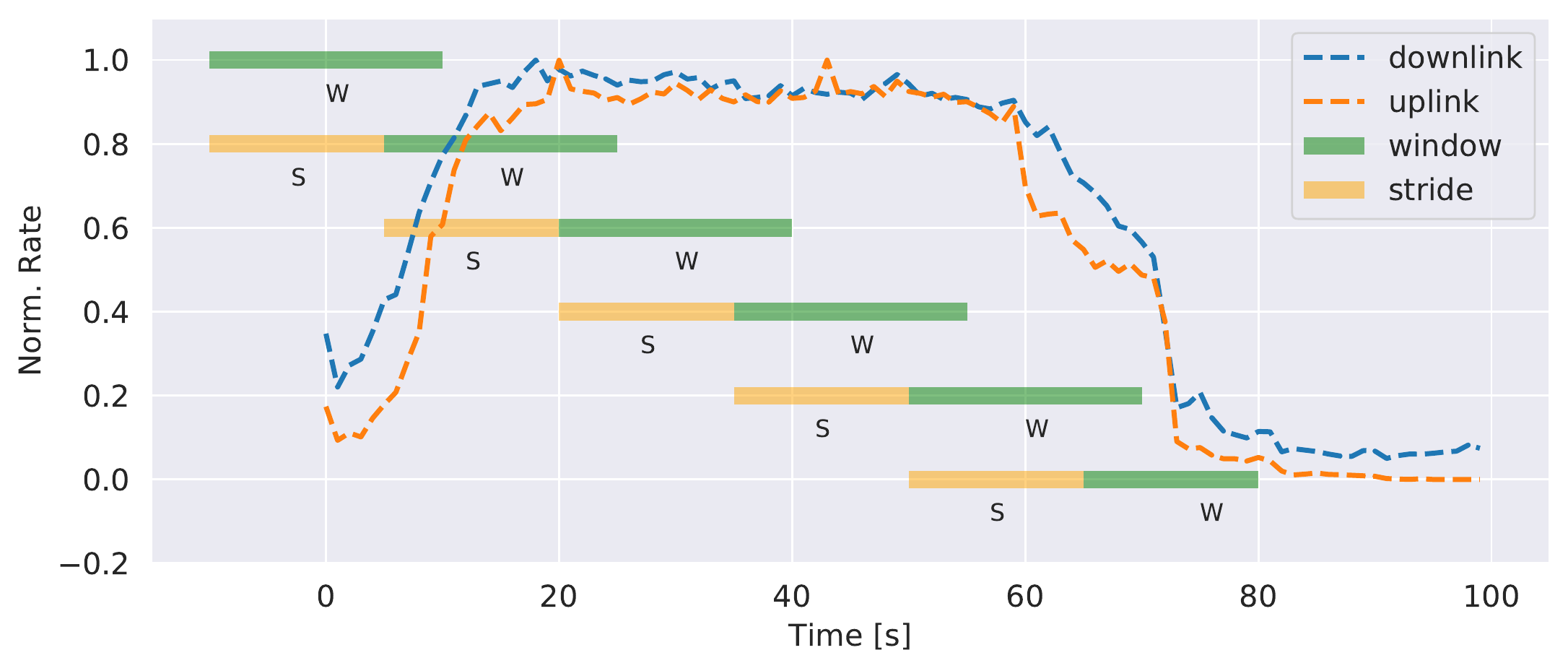}
	\caption{Sliding window of $20$s length and $15$s stride, applied to a sample \mbox{video-streaming} session.}
	\label{fig:slide_win}
\end{figure}

To ensure the applicability of our classifiers to real world (asynchronous) cases, we account for \textit{asynchronous} sessions, entailing that the classification algorithm has no knowledge about the instants where the sessions begin and end. Specifically, each session is split using a sliding window of length $W$~seconds, moved rightwards from the beginning of the session with a stride of $S$~seconds, see Fig.~\ref{fig:slide_win}. The split sessions (asynchronous sessions), of $W$ seconds each, represent the input data to our classification algorithms. Note that $W$ and $S$ are \mbox{hyper-parameters} of the proposed classification frameworks. 

\subsection{Sessions Correlation over Time}

\begin{figure}[t!]
	\centering
	\includegraphics[width=.9\columnwidth]
	{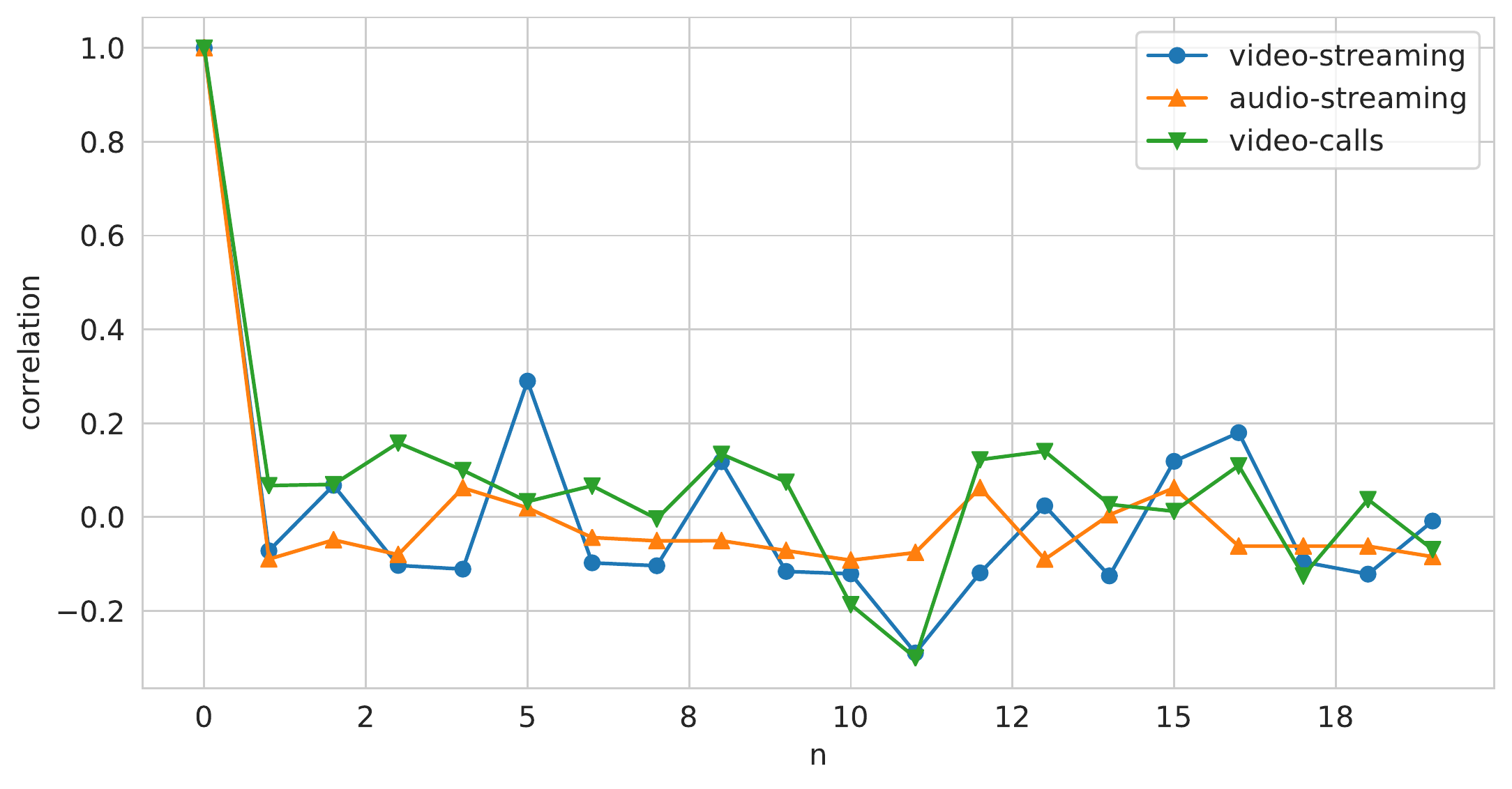}
	\caption{
	Pearson correlation between the initial and the following sessions running in the controlled UE.
	}
	\label{fig:correlation}
\end{figure}

As a sanity check, we verify the soundness of the watermarking strategy: our aim is to understand whether the transmission of user data in the form of duty-cycled patterns may affect the way in which the eNodeB handles the communication from our terminal, e.g., through some advanced channel reservation mechanism. In that case, in fact, our watermarking strategy would be of little use, as it would introduce scheduling artifacts that do not occur in real life conditions. To verify this, we evaluated the Pearson correlation between the initial session (i.e., when the \ac{UE} connects to the LTE PDCCH for the first time and it is assigned a new \mbox{C-\ac{RNTI}}) and the following ones. Fig.~\ref{fig:correlation} shows that, for each of the three services, the correlation is high only when we compare the first session with itself ($n=0$). Instead, low values are observed between the first session and the following ones ($n>1$), indicating that the behavior of the eNodeB scheduler is not affected by the repetitive actions (i.e., the \mbox{duty-cycled} activity) performed at the \ac{UE} side. 

\section{Classification Problem}
\label{sec:sl}

\subsection{Problem Definition}
\label{sec:problem}


Let $M$ be the number of windowed-sessions obtained through the data preparation procedure of Section~\ref{sec:labeled_dataset}, $W$ is the window size, and $D = 2$ is the number of communication directions (downlink and uplink). We define $\bm{X}$ the input dataset tensor with size $M \times W \times D$, where the \mbox{$m$-th} row vector $\bm{x}_m$ contains the trace associated with $W$ \ac{TBS} samples per session for both downlink and uplink directions ($D=2$).

A classifier estimates a function $c:\bm{X} \rightarrow \bm{Y}$, where the output matrix $\bm{Y}$ has size $M \times K$, with $K$ representing the number of classes. The row vector \mbox{$\bm{y_m}=c(\bm{x_m})=[y_{m1},\dots, y_{mK}]$} contains the probabilities that session $m$ belongs to each of the $K$ classes, with $\sum_{k} y_{mk} =1$. The final output of the classifier is class $k^\star$, where $k^\star = \arg\!\max_k (y_{mk})$. The following classification objectives are addressed:
\begin{enumerate}
	\item[O1)] \textbf{Service identification:} to classify the collected sessions into $K = 3$ classes, namely, {\it audio streaming}, {\it video streaming} and {\it video calls};
	\item[O2)] \textbf{App identification:} to identify which app is run at the \ac{UE}. In this case, the number of output classes is $K=6$, namely, \textit{Spotify}, \textit{Google Music}, \textit{YouTube}, \textit{Vimeo}, \textit{Skype} and \textit{WhatsApp Messenger}.
\end{enumerate}


Next, we present the considered classification algorithms, grouping them into two categories: those based on artificial neural networks and those based on standard machine learning techniques (referred to here as benchmark classifiers). 

\subsection{Deep Neural Networks}
\label{sec:nn}

Next, we describe how we tailored three neural network architectures to solve the above traffic classification problem, namely,  \ac{MLP}, Recurrent Neural Networks (RNNs) and Convolutional Neural Networks (CNNs). 


\subsubsection{Multilayer Perceptron}
\label{sec:nn_mlp}

A multilayer perceptron is a feedforward and \mbox{fully-connected} neural network architecture. The term \mbox{``feed-forward''} refers to the fact that the information flows in one direction, from the input to the output layer. An MLP is composed of, at least, three layers of nodes: an input, a hidden and an output layer. A directed graph connects the input with the output layer and each neuron in the graph uses a \mbox{non-linear} activation function to produce its output. Links are weighted and the backpropagation algorithm is utilized to train the network in a supervised fashion, i.e., to find the set of network weights that minimize a certain error function, given an input set of examples and the corresponding labels. For further details, see~\cite{bishop2006}.

The \ac{MLP} that we use for mobile traffic classification has \emph{three} fully connected layers. The first layer ${MLP}_1$ contains $N_{{MLP}_1}= 128$ neurons, the second layer ${MLP}_2$ has $N_{{MLP}_2}= 64$ neurons and the third layer $MLP_3$ is fully connected, with $K$ neurons and a softmax activation function to produce the final output. The output of ${MLP}_3$ is the class probability vector $\bm{y_m}$.

%

All neurons in layers $MLP_1$ and $MLP_2$ use a leaky version of the \ac{ReLU} ({\it leaky \ac{ReLU}}) activation function. Leaky \acp{ReLU} help solve the vanishing gradient problem, i.e., the fact that the error gradients that are backpropagated during the training of the network weights may become very small (zero in the worst case), preventing the correct (gradient based) adaptation of the weights. To prevent this from happening, leaky \acp{ReLU} have a small negative slope for negative values of their argument~\cite{maas2013rectifier}.
To train the presented MLP architecture, we use the {\it RMSprop} gradient descent algorithm~\cite{tieleman2012lecture}, by minimizing the {\it categorical \mbox{cross-entropy} loss function} $L(\bm{w})$, defined as~\cite{bishop2006} 
\begin{equation}
\label{eq:cross}
L(\bm{w}) = - \sum_{\bm{x}_m \in \bm{B}} \sum_{k=1}^{K} t_k(\bm{x}_m) \log(y_{mk}(\bm{w}, \bm{x}_m)).
\end{equation}
where $\bm{t(x)_m} = [t_1(\bm{x}_m), \dots, t_k(\bm{x}_m)]$ contains the class labels associated with the input trace $\bm{x}_m$, i.e., $t_k=1$ if $\bm{x}_m$ belongs to class $k$ and $t_k = 0$ otherwise (\mbox{$1$-of-$K$} coding scheme). Vector $\bm{w}$ contains the \ac{MLP} weights and $y_{mk}(\bm{w}, \bm{x}_m)$ is the MLP output obtained for input $\bm{x}_m$. Eq.~(\ref{eq:cross}) is iteratively minimized using the training examples in the batch set $\bm{B}\subset \bm{X}$, where $\bm{B}$ is changed at every iteration so as to span the entire input set $\bm{X}$.

\subsubsection{Recurrent Neural Networks}
\label{sec:nn_rnn}

Recurrent Neural Networks (RNNs) have been conceived to extract features from temporal (and correlated) data sequences. \ac{LSTM} networks are a particular type of \ac{RNN}, introduced in \cite{Hochreiter1997}. They are capable of tracking \mbox{long-term} dependencies into the input time series, while solving the \mbox{vanishing-gradient} problem that affects standard \acp{RNN}~\cite{gers1999learning}.

The capability of learning \mbox{long-term} dependencies is due to the structure of the \ac{LSTM} cells, which incorporates gates that regulate the learning process. The neurons in the hidden layers of an \ac{LSTM} are \acp{MC}. A \ac{MC} has the ability to retain or forget information about past input values (whose effect is stored into the cell states) by using structures called {\it gates}, which consist of a cascade of a neuron with sigmoidal activation function and a pointwise multiplication block. Thanks to this architecture, the output of each memory cell possibly depends on the entire sequence of past states, making \acp{LSTM} suitable for processing time series with long time dependencies~\cite{Hochreiter1997}. The input gate of a memory cell is a neuron with sigmoidal activation function ($\sigma$). Its output determines the fraction of the MC input that is fed to the cell state block. Similarly, the forget gate processes the information that is recurrently fed back into the cell state block. The output gate, instead, determines the fraction of the cell state output that is to be used as the output of the \ac{MC}, at each time step. Gate neurons usually have sigmoidal activation functions ($\sigma$), while the hyperbolic tangent ($\tanh$) activation function is usually adopted to process the input and for the cell state. All the internal connections of the \ac{MC} have unitary weight~\cite{Hochreiter1997}.

\begin{figure}[t]	
	\centering
	\includegraphics[width=.8\columnwidth]{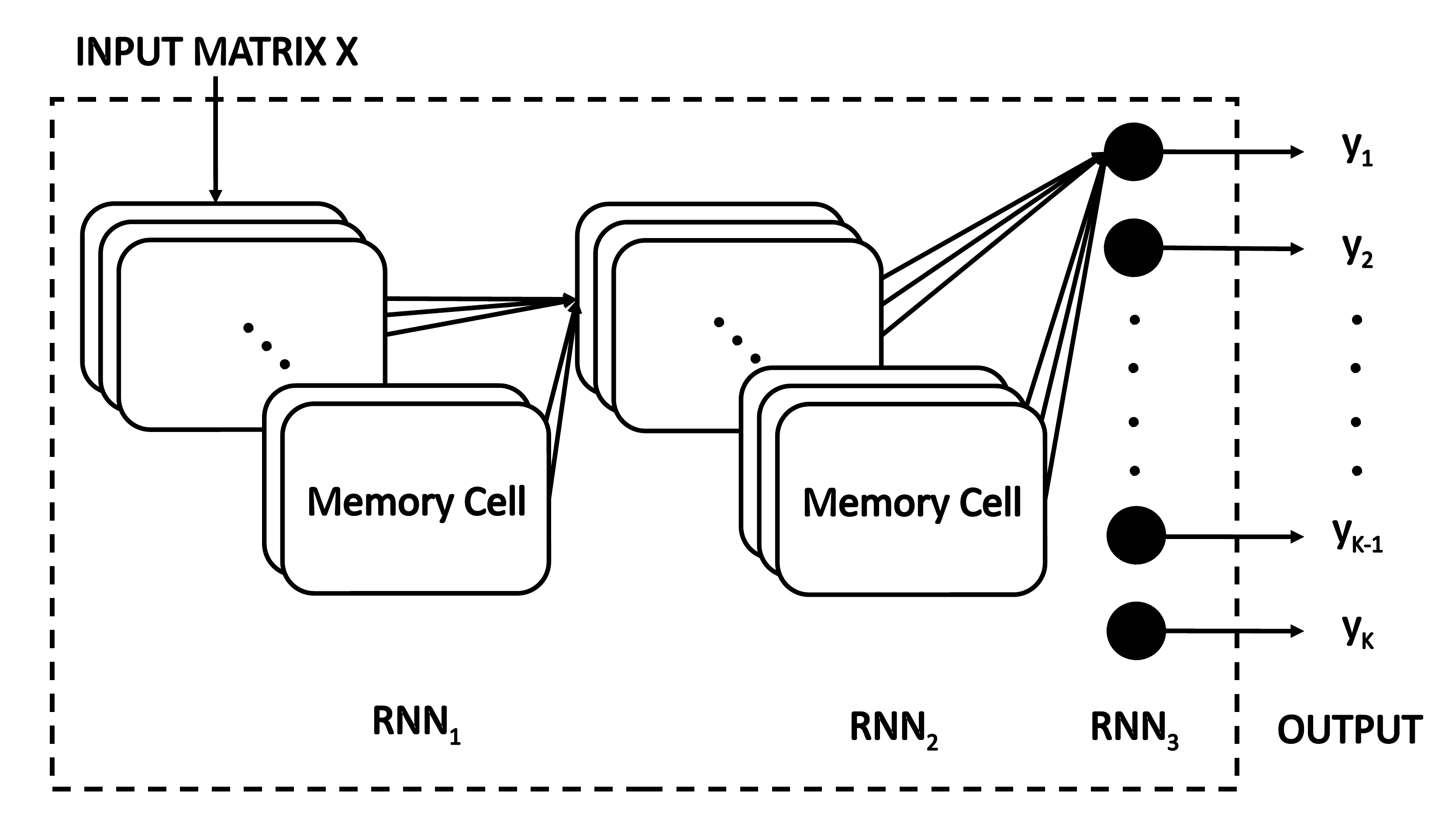}
	\caption{\ac{RNN} architecture.}
	\label{fig:lstm_nn}	
\end{figure}

The proposed RNN based traffic classification architecture is shown in Fig.~\ref{fig:lstm_nn}. In our design, we consider three stacked layers combining two LSTM layers and a final fully connected output layer. The first and the second layer (respectively ${RNN}_1$ and ${RNN}_2$) have \mbox{$N_{{RNN}_1}=N_{{RNN}_2}= 180$} memory cells. The fully connected layer ${RNN}_3$ uses the softmax activation function and its output consists of the class probability estimates, as described in Section~\ref{sec:nn_mlp}.


\subsubsection{Convolutional Neural Networks}
\label{sec:nn_cnn}

\begin{figure}[t]	
	\centering
	\includegraphics[width=.8\columnwidth]{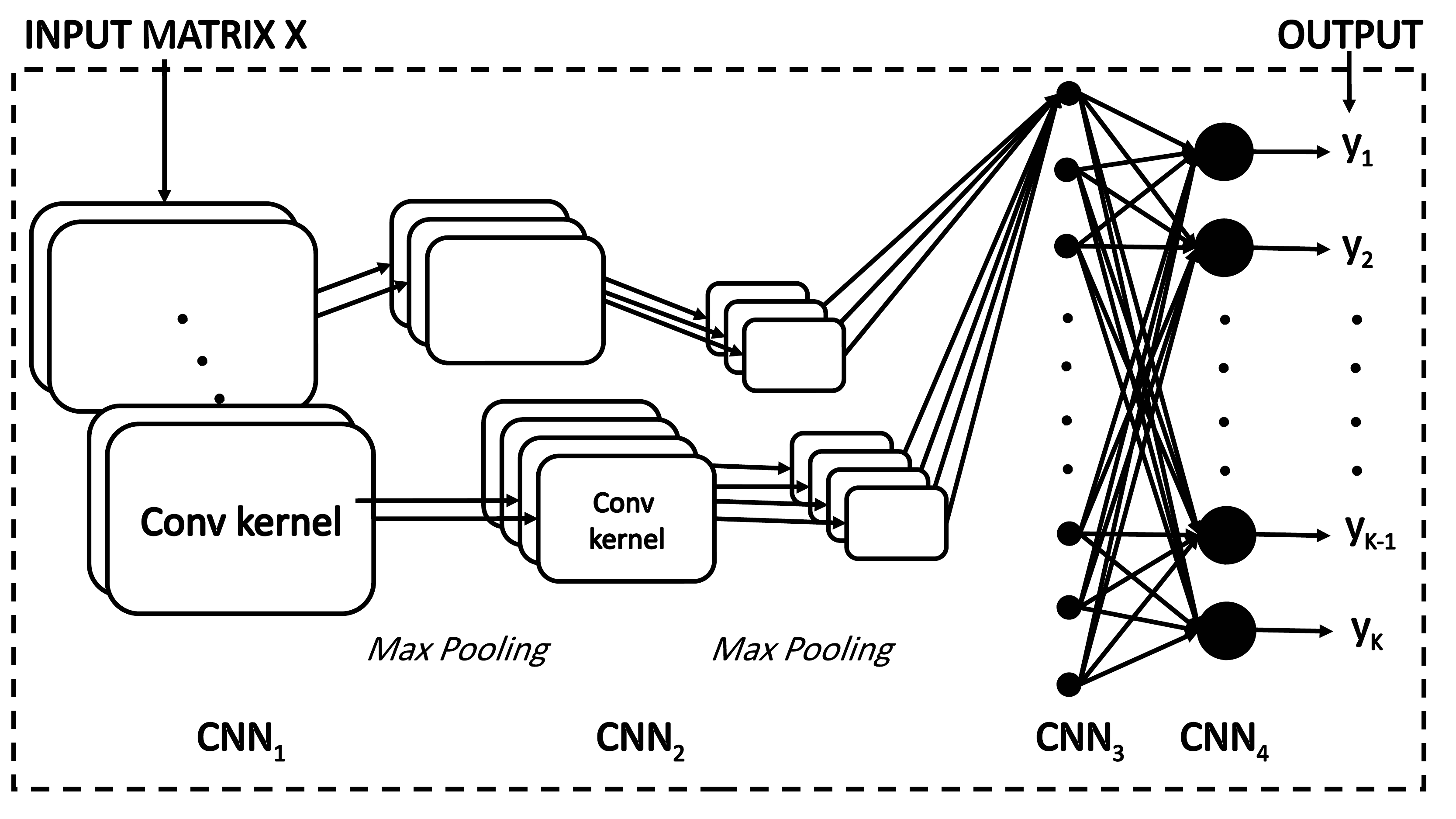}
	\caption{\ac{CNN} architecture.}
	\label{fig:cnn}	
\end{figure}

Convolutional Neural Networks (CNNs) are \mbox{feed-forward} deep neural networks differing from fully connected MLP for the presence of one or more convolutional layers. At each convolutional layer, a number of kernels is used. Each kernel is composed of a number of weights and is convolved across the entire input signal. Note that the kernel acts as a filter whose weights are \mbox{re-used} ({\it shared weights}) across the entire input and this makes the network connectivity structure sparse, i.e., a small set of parameters (the kernel weights) suffices to map the input into the output. This leads to a considerably reduced computational complexity with respect to fully connected feed forward neural
networks, and to a smaller memory footprint. For more details the reader is referred to~\cite{goodfellow2016deep}. 

\acp{CNN} have been proven to be excellent feature extractors for images and inertial signals~\cite{gadaleta2016idnet} and here we show their effectiveness for the classification of mobile traffic data. The \ac{CNN} architecture that we designed to this purpose is shown in Fig.~\ref{fig:cnn}. It has two main parts: the first four layers perform convolutions and max pooling in cascade, the last two are fully connected layers. 
The first convolutional layer $CNN_1$ uses one dimensional kernels ($ 1\times 5$ samples) performing a first filtering of the input and processing each input vector (rows of $\bm{X}$) separately. The activation functions are linear and the number of convolutional kernels is $N_{{CNN}_1}=32$. The second convolutional layer, $CNN_2$, uses one dimensional kernels ($1 \times 5$ samples) with \mbox{non-linear} hyperbolic tangents as activation functions, and the number of convolutional kernels is $N_{{CNN}_2}=64$.
Max pooling is separately applied to the outputs of $CNN_1$ and $CNN_2$ to reduce their dimensionality and increase the spatial invariance of features~\cite{gadaleta2016idnet}. In both cases, a \mbox{one-dimensional} pooling with a kernel of size $1 \times 3$ is performed. A third fully connected layer, $CNN_3$, performs dimensionality reduction and has $N_{{CNN}_3}=32$ neurons with Leaky \ac{ReLU} activation functions. This layer is used in place of a further convolutional layer to reduce the computation time, with a negligible loss in accuracy. The last (output) layer $CNN_4$ is fully connected with softmax activation functions, and returns the class probability estimates, see Sections~\ref{sec:nn_mlp}.

\subsection{Benchmark classifiers}
\label{sec:benchmark}

Other standard classification schemes have been tailored to the considered tasks O1 and O2. The selected algorithms are: \textit{Linear Logistic Regression}, \textit{$K$-Nearest Neighbours} and \textit{Linear SVM}, as examples of linear classifiers; \textit{Random Forest}, as an ensemble learning method, and \textit{Gaussian Processes} as an instance of Bayesian approaches. 
The implementations of Linear Logistic Regression, \mbox{$K$-Nearest Neighbours} and Linear SVM are based on~\cite{fan2008liblinear},~\cite{altman1992introduction} and~\cite{wu2007robust}, respectively. The Random Forest implementation is based on~\cite{breiman2001random}, whereas for the classifier based on Gaussian Processes we refer to~\cite{rasmussen2004gaussian}. Configuration parameters and implementation details for the benchmark classifiers are provided in Table~\ref{table_params}.


\begin{table}[t]
\centering
\caption{Configuration parameters for the benchmark classifiers.}
\ra{0.5}
\scriptsize
\begin{tabular}{p{1.2cm}p{2.4cm}p{3.1cm}}\toprule
\textbf{Algorithm}&\hspace{0.2cm}\textbf{Parameters}& \hspace{0.2cm}\textbf{Note - Reference} \\
\midrule
\rowcolor{black!5} Linear SVM & \begin{tabular}{p{2.4cm}}
\tabitem \textit{penalty} = L2 \\
\tabitem \textit{loss} = Hinge Loss \\
\tabitem $c = 0.025$
\end{tabular} & \begin{tabular}{p{3.1cm}}
\tabitem $c$: penalty parameter for the error term\\
\tabitem extended to multi-class with one-vs-rest~\cite{fan2008liblinear}
\end{tabular}\\
\midrule
\rowcolor{black!0} Logistic Regressor& \begin{tabular}{p{2.4cm}}
\tabitem \textit{penalty} = L2 \\
\tabitem $c = 1$
\end{tabular} & \begin{tabular}{p{3.1cm}}
\noindent \tabitem $c$: inverse of the regularization strength\\
\tabitem extended to multi-class with one-vs-rest~\cite{fan2008liblinear}
\end{tabular}\\
\midrule
\rowcolor{black!5} Nearest Neighbours & \begin{tabular}{p{2.4cm}}
\tabitem $K = 3$\\
\tabitem $p = 2$\\
\tabitem \textit{metric} = Minkowski   
\end{tabular} & \begin{tabular}{p{3.1cm}}
\tabitem $K$: number of neighbors for queries\\ 
\tabitem $p$: distance metric parameter\\
\tabitem $p = 2$ amounts to using the Euclidean distance~\cite{altman1992introduction}
\end{tabular}\\
\midrule
\rowcolor{black!0 } Random Forest & \begin{tabular}{p{2.4cm}}
\tabitem \textit{n. estimators} $= 10$\\
\tabitem \textit{max depth} $= 5$\\
\tabitem \textit{criterion} $=$ entropy\\
\end{tabular} & \begin{tabular}{p{3.1cm}}
\tabitem \textit{n. estimators}: number of trees in the forest\\ 
\tabitem  \textit{max depth}: maximum depth of a tree\\
\tabitem  \textit{criterion}: function to measure the quality of a split of subsets~\cite{breiman2001random}\\
\end{tabular}\\
\midrule	
\rowcolor{black!5} Gaussian Processes & \begin{tabular}{p{2.4cm}}
\tabitem \textit{kernel} $=$ RBF\\
\tabitem $\sigma =$ Logistic func.\\
\tabitem \textit{approx.} $=$ Laplacian
\end{tabular} & \begin{tabular}{p{3.1cm}}
\tabitem Radial Basis Function (RBF) used as kernel\\ 
\tabitem $\sigma$ is the sigmoid function used to ``squash'' the nuisance function\\ 							
\tabitem Laplacian method used to approximate the non Gaussian Posterior~\cite{rasmussen2004gaussian}	
\end{tabular}\\
\bottomrule
\end{tabular}
\label{table_params}
\end{table}

\section{Supervised Training and Comparison of Traffic Classification Algorithms}
\label{sec:results}

\begin{table*}[t]
\centering
\ra{0.5}
\footnotesize
\begin{tabular}{*{8}{c}}
\textbf{Algorithm}&\multicolumn{1}{c}{\textbf{Accuracy $\%$}} & \multicolumn{1}{r}{\textbf{Precision}} & \textbf{Recall} & \textbf{F-Score} & \textbf{\# Parameters}  & \textbf{Acc Async $\%$} & \textbf{Diff $\%$} \\
\midrule
\rowcolor{black!5} Linear SVM & 81.23 & 0.811 & 0.812 & 0.805  & 36726 & 68.41 & -12.8\\
\rowcolor{black!0}Logistic Regressor & 81.61 & 0.806 &0.816& 0.809   &  486 & 65.72 & -15.9 \\
\rowcolor{black!5}Nearest Neighbours   & 84.51 & 0.843 &0.845& 0.841   &   36720 & 79.65 & -4.9\\
\rowcolor{black!0}Random Forest  & 83.52 & 0.821& 0.835& 0.827  &   41310 & 70.21 & -13.3\\
\rowcolor{black!5}Gaussian Processes  & 87.43 & 0.874 & 0.871  & 0.871   &   146720 & 81.21 & -6.2\\
\midrule
\textbf{Neural Networks} &  &   &    &    &  \\
\midrule
\rowcolor{black!5}MLP & 90.04 & 0.900 & 0.900 & 0.900  &  19014 & 84.61 & -5.4 \\
\rowcolor{black!0} RNN  &  96.57 & 0.967  & 0.968   & 0.968  & 392046  & 92.93 & -3.6\\
\rowcolor{black!5}CNN & 97.77 & 0.978 & 0.976  & 0.977   & 25062 & 93.20 & -4.5\\
\bottomrule

\end{tabular}

\caption{Classifiers comparison for the app identification task.}
\label{table_app}
\end{table*}

\begin{table*}[t]
\centering
\ra{0.5}
\footnotesize
\begin{tabular}{*{8}{c}}
\textbf{Algorithm}&\multicolumn{1}{c}{\textbf{Accuracy $\%$}} & \multicolumn{1}{r}{\textbf{Precision}} & \textbf{Recall} & \textbf{F-Score} & \textbf{\# Parameters} & \textbf{Acc Async $\%$} & \textbf{Diff $\%$} \\
\midrule
\rowcolor{black!5} Linear SVM & 90.80 & 0.908& 0.908& 0.907  &  19843 & 79.61  & -11.2\\
\rowcolor{black!0}Logistic Regressor & 90.42 & 0.904& 0.904& 0.904   & 243 & 81.11 & -9.3\\
\rowcolor{black!5}Nearest Neighbours   & 92.76 & 0.925& 0.925& 0.925 &  19840 & 83.45 & -9.3\\
\rowcolor{black!0}Random Forest  & 91.57 & 0.915& 0.915& 0.915 &  22320 & 84.25 & -7.3\\
\rowcolor{black!5}Gaussian Processes  &  93.21 & 0.932 & 0.928 & 0.929 &  73360 & 82.51 & -10.7\\
\midrule
\textbf{Neural Networks} &  &   &    &    & \\
\midrule
\rowcolor{black!5}MLP &  94.31 & 0.943 & 0.939 & 0.942  & 18819 & 93.38 & -0.9 \\
\rowcolor{black!0} RNN  &  98.21 & 0.981 & 0.982 & 0.981 & 391503 & 95.38 & -2.8\\
\rowcolor{black!5}CNN & 98.87 & 0.986 & 0.988 & 0.988  & 24963 & 95.40 & -3.5\\
\bottomrule
\end{tabular}
\caption{Classifiers comparison for the service identification task.}
\label{table_service}

\end{table*}


The performance tests have been carried out using an Intel core i7 machine, with $32$ GB of RAM and an NVIDIA GTX 980 GPU card. We divided the dataset, featuring $10,000$ labeled DCI sessions, into training and validation sets with a split ratio of \mbox{$70\%$ - $30\%$}. These sets are balanced, as they contain the same percentage of traces for all classes. The classification algorithms have been implemented in Python. We have used \verb|keras| library on top of Tensorflow backend for the implementation of deep NNs. For the benchmark classifiers, we used the popular \verb|sklearn| library.



\subsection{Performance Metrics}
\label{sec:metrics}

The classification performance is assessed through the following metrics:
\begin{enumerate}
\item \textbf{Accuracy}: defined as the ratio between the number of correctly classified sessions to the total number of sessions.
\item \textbf{Precision} $P$: defined, for each class, as the ratio between true positives $T_p$ and the sum between true positives and false positives $F_p$,
\begin{equation}
P = \frac{T_p}{T_p+F_p},
\end{equation}

\item \textbf{Recall} $R$: defined, for each class, as the ratio between the true positives $T_p$ and the sum of true positives and false negatives $F_n$,
\begin{equation}
R = \frac{T_p}{T_p+F_n}.
\end{equation}

\item \textbf{F-Score} $F$ is defined as the harmonic mean of precision $P$ and recall $R$,
\begin{equation}
\displaystyle F = \left(\frac{\frac{1}{P}+\frac{1}{R}}{2}\right)^{-1}=2\frac{RP}{R+P}.
\end{equation}
\end{enumerate}
Note that the definition of precision and recall only applies to classification tasks with one class. However, tasks O1 and O2 both have a number of classes $K >2$, namely, $K=\{3,6\}$ for app and service identification, respectively. Thus, precision and recall are separately calculated for all the $K$ classes, and their average is shown in the following numerical results.


\subsection{Comparison of Classification Algorithms}
\label{sec:perf_ev}
\subsubsection{Accuracy and Algorithm Training}

\begin{figure*}[t]
	\centering
	\begin{subfigure}[t]{.8\columnwidth}
		\centering
		\includegraphics[width=0.98\columnwidth]{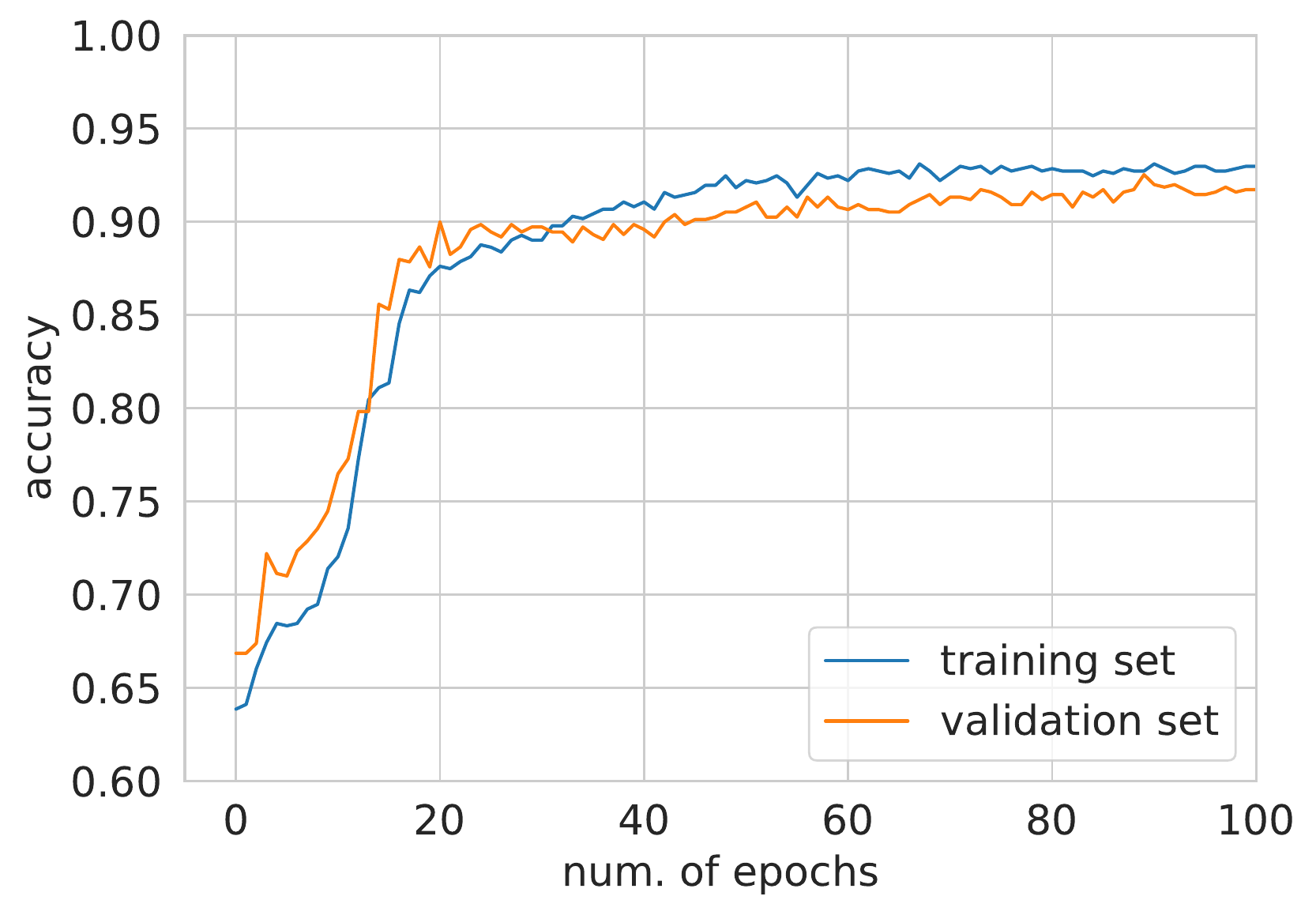}
		\caption{RNN.}
		\label{fig:train_test_RNN}	
	\end{subfigure}
	\begin{subfigure}[t]{.8\columnwidth}
		\centering
		\includegraphics[width=0.98\columnwidth]{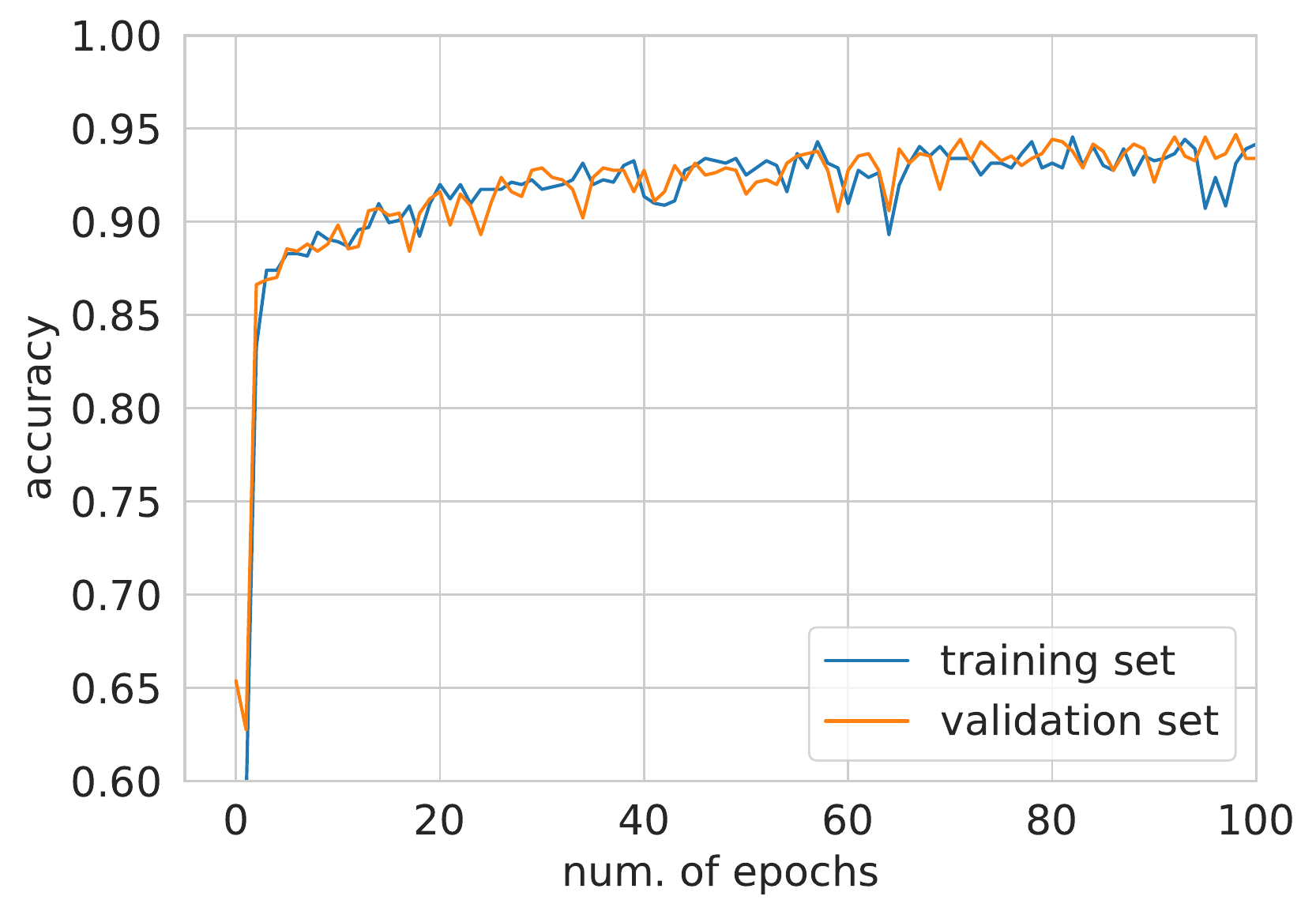}
		\caption{CNN.}
		\label{fig:train_test_CNN}	
	\end{subfigure}
	\centering
	\caption{Accuracy {\it vs} number of epochs for training and validation sets for the app identification task.}
	\label{fig:train_test}
\end{figure*}

Tables~\ref{table_app} and~\ref{table_service} summarize the obtained performance metrics for the deep \acp{NN} and the benchmark classifiers for app and service identification, respectively. First, we focus on synchronous sessions results. In general, better performance is achieved through deep \acp{NN} ($+13.8\%$ on app identification, $+8.7\%$ on service classification). Moreover, we observe a significant performance gap between the service and app classification tasks, due to the higher number of classes of the latter: the performance gap is higher than $8\%$ for the benchmark classifiers and ranges from $2$ to $6\%$ for NNs. Furthermore, \ac{RNN} and \ac{CNN} architectures achieve an accuracy of about $99\%$ for the service identification task (O1) and higher than $95\%$ for the app identification task (O2). The algorithm based on Gaussian Processes performs the best among the benchmark classifiers. In general, the higher the complexity (i.e., the number of parameters, and also hidden layers for NNs), the higher the performance. The only exception to this is provided by \acp{CNN}, which present the highest accuracy but use a small number of parameters. This fact confirms the high efficiency of convolutions in processing high amount of data with complex temporal structure, and the effectiveness of parameter sharing. \acp{CNN} require only $6\%$ of the variables used up by \acp{RNN}, achieving a better accuracy. This also translates into a faster training: in Fig.~\ref{fig:train_test}, we show the accuracy as a function of the number of epochs for training and validation sets for RNNs and CNNs. The number of epochs required by the \acp{CNN} to reach an accuracy higher than $90\%$ is fewer than $20$ (Fig.~\ref{fig:train_test_CNN}), whereas for RNNs convergence is achieved only after $30$ epochs (Fig.~\ref{fig:train_test_RNN}).

\subsubsection{Confusion Matrix}

\begin{figure*}[h]
	\centering
	\begin{subfigure}[t]{.8\columnwidth}
		\centering
		\includegraphics[width=0.98\columnwidth]{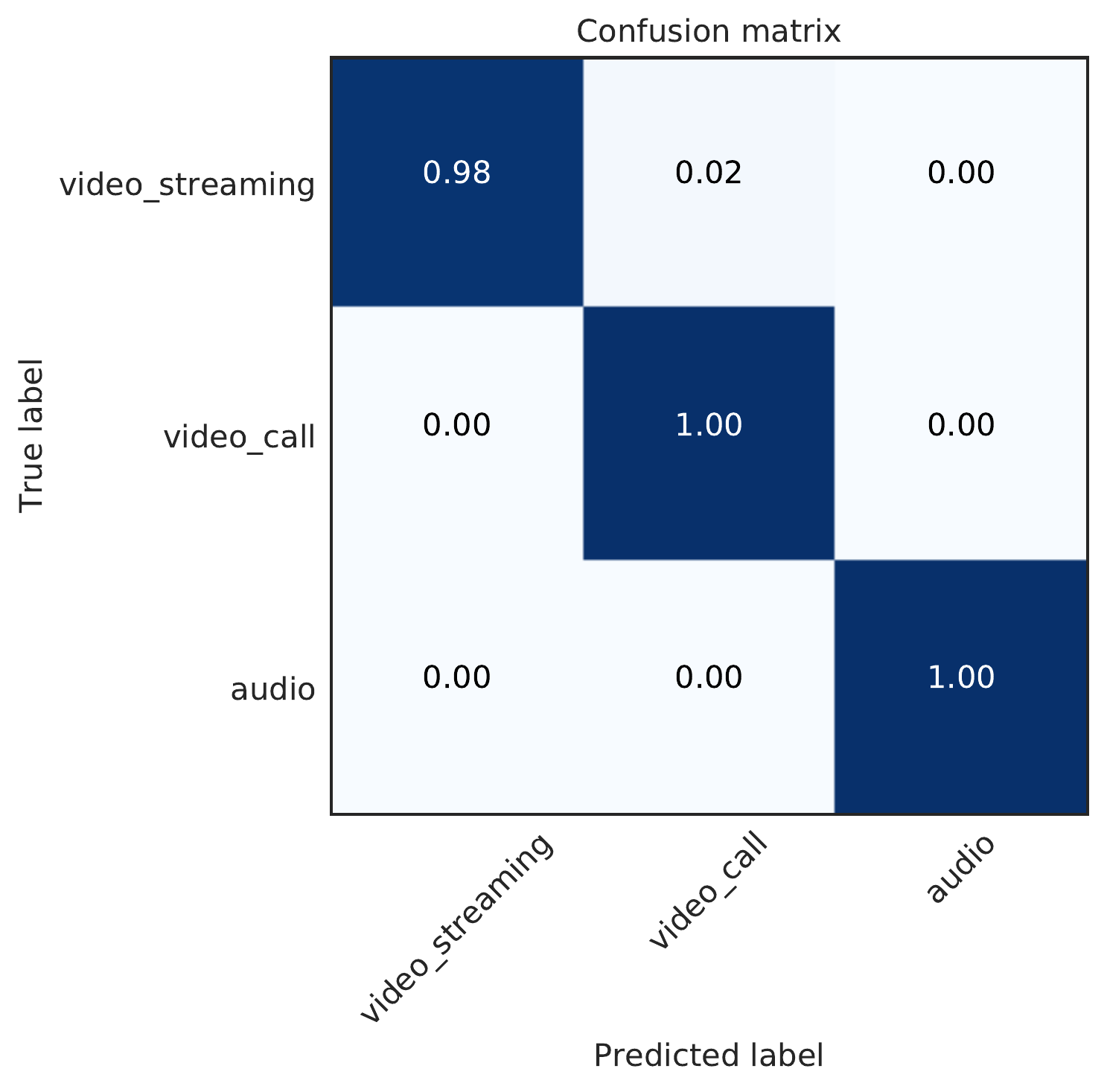}
		\caption{Confusion matrix for service identification.}
		\label{fig:conf_service}	
	\end{subfigure}
	\begin{subfigure}[t]{.8\columnwidth}
		\centering
		\includegraphics[width=0.98\columnwidth]{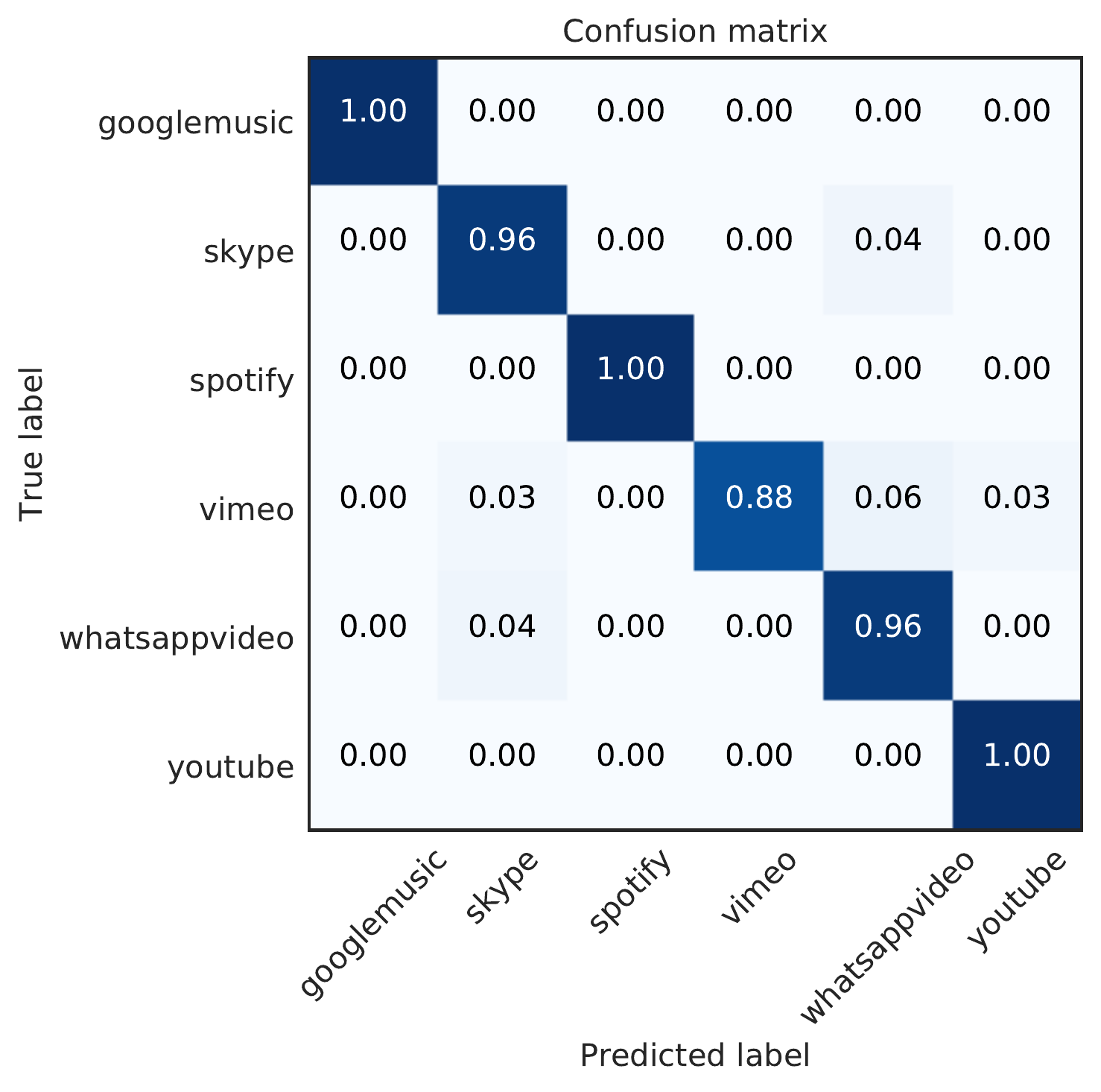}
		\caption{Confusion matrix for app identification.}
		\label{fig:conf_app}	
	\end{subfigure}
	\centering
	\caption{Confusion matrices for the \ac{CNN} algorithm.}
	\label{fig:conf_matrices}
\end{figure*}

A deeper look at the performance of CNNs is provided by the confusion matrices of Fig.~\ref{fig:conf_matrices}, whose rows and columns respectively represent true and predicted labels, and all values are normalized between $0$ and $1$. For the service classification task (Fig.~\ref{fig:conf_service}), CNNs only misclassify the video streaming sessions: $2\%$ of those are labeled as video calls. For the app identification task (Fig.~\ref{fig:conf_app}), errors ($4\%$) mainly occur for Skype and WhatsApp videocalls. These errors are understandable, as these are both interactive real-time video applications and, as such, their traffic patterns bear similarities. The lowest performance is found for Vimeo traces, for which $88\%$ of the sessions are correctly classified. Here our CNN-based classifier confuses them with the other video applications for both streaming service (Youtube - $3\%$) and real-time calling (WhatsApp and Skype - $6\%$ and $3\%$, repectively).

\subsubsection{Asynchronous Sessions Results}

As shown in the last two columns of Tables~\ref{table_app} and~\ref{table_service}, the algorithms' accuracy is affected by asynchronous sessions. As expected, we observe a general decrease in the accuracy for all the algorithms ($-6.0\%$ for service identification, $-7.7\%$ for app identification, on average). However, for both classification problems, neural network-based approaches are more robust to the asynchronous case, showing a performance degradation of $-4.3$\%, while the degradation of standard algorithms is $-8.4$\%.

\subsubsection{Impact of Different Window Sizes}
\begin{figure*}[t]
	\centering
	\begin{subfigure}[t]{.8\columnwidth}
		\centering
		\includegraphics[width=0.98\columnwidth]{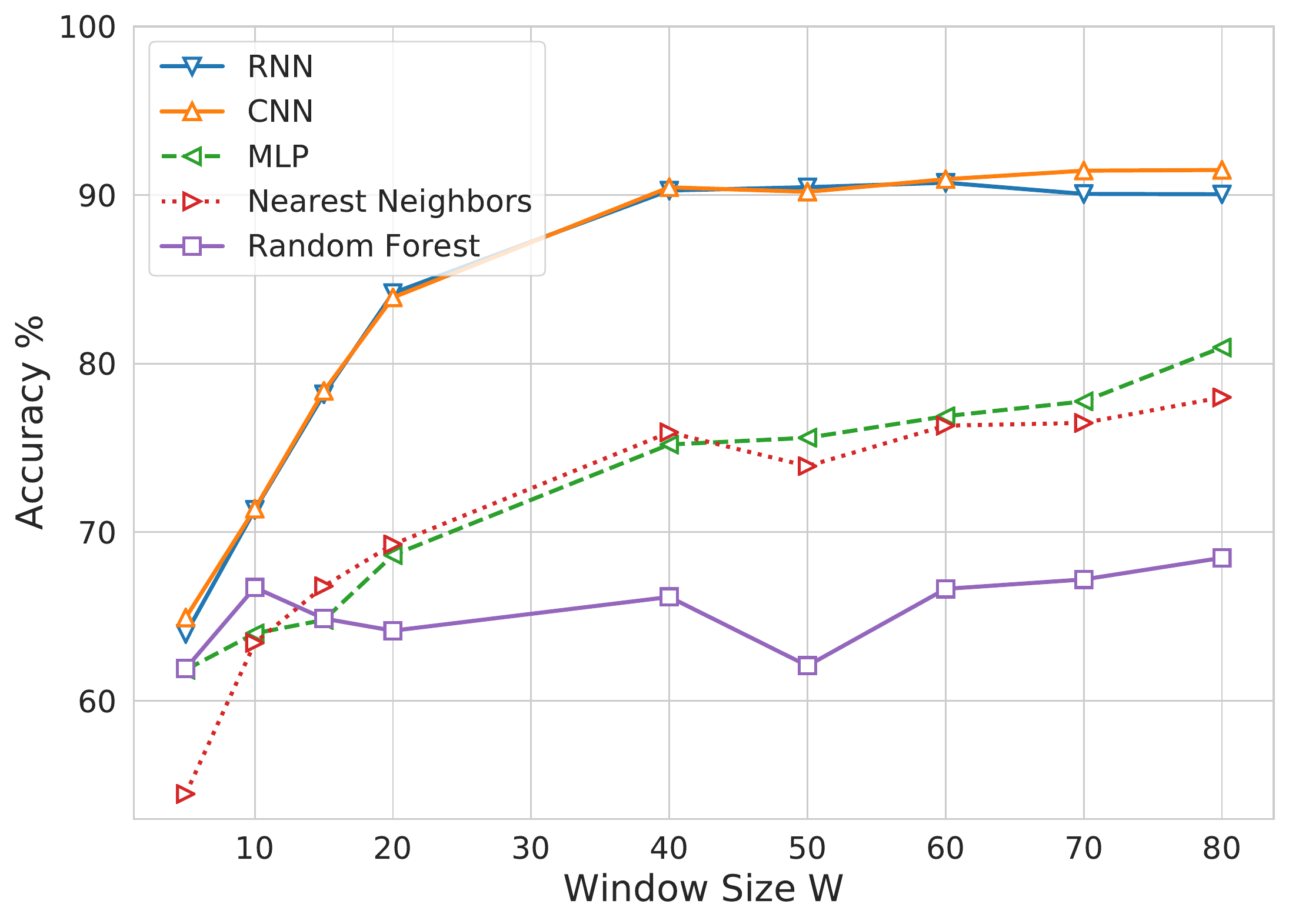}
		\caption{Accuracy for app identification.}
		\label{fig:bench_vs_W_app}	
	\end{subfigure}
	\begin{subfigure}[t]{.8\columnwidth}
		\centering
		\includegraphics[width=0.98\columnwidth]{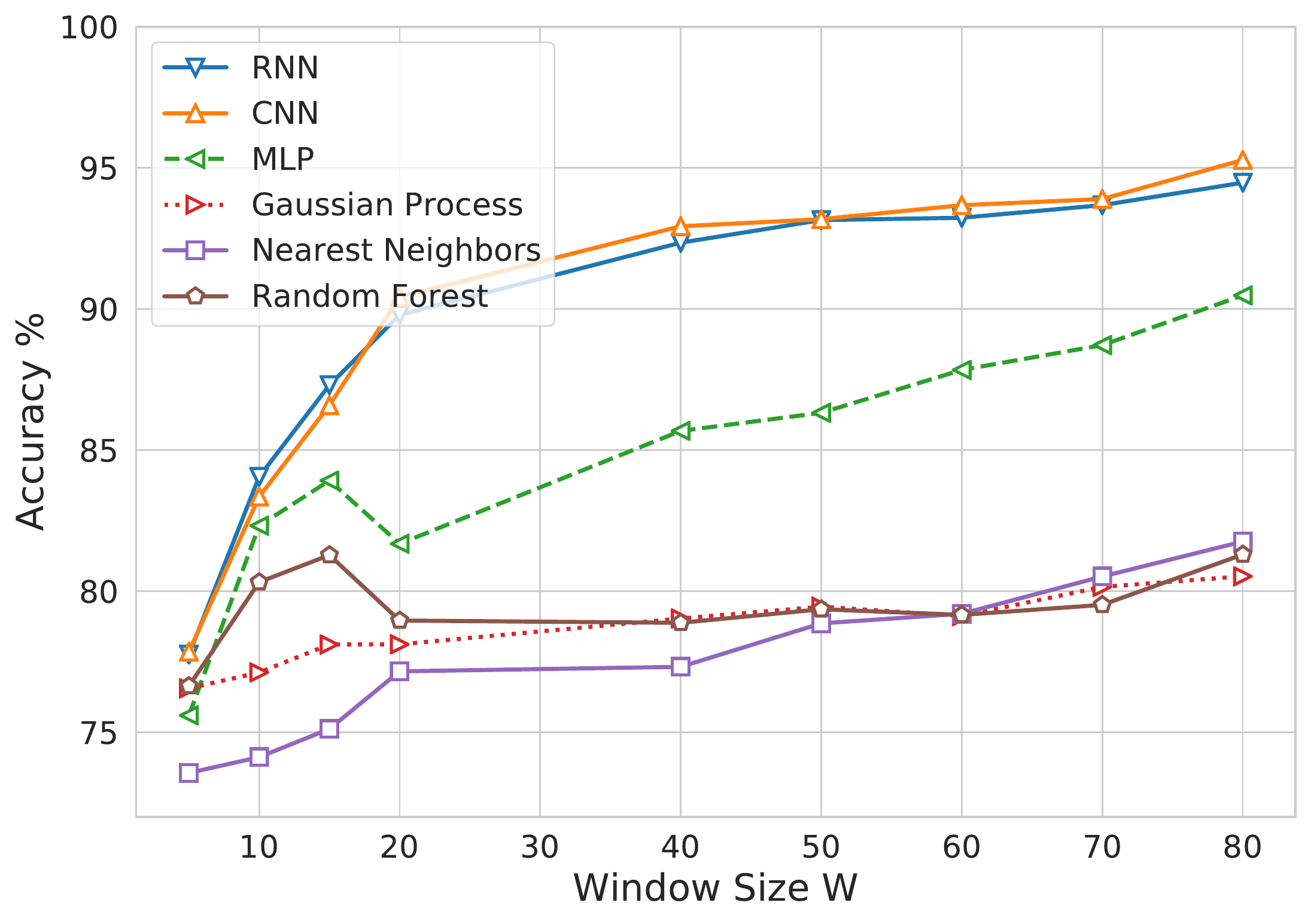}
		\caption{Accuracy vs Window Size $W$ for service identification.}
		\label{fig:bench_vs_W_service}	
	\end{subfigure}
	\centering
	\caption{
	Accuracy {\it vs} window lengths for app and service classification tasks.}
	\label{fig:accuracy_vs_W}
\end{figure*}

Fig.~\ref{fig:accuracy_vs_W} shows the classification accuracy as a function of the window size, $W$. For the app identification task, $40$~seconds suffice for \acp{CNN} and \acp{RNN} to reach accuracies higher than $90\%$, with negligible additional improvements for longer observation periods. Periods shorter than $40$ seconds provide less accurate results. Similar trends are observed for the service classification task. However, in this case after $20$ seconds the accuracy of \acp{CNN} and \acp{RNN} is already higher than $90\%$, due to the smaller number of classes. In summary, the ability of \acp{CNN} and \acp{RNN} to extract representative statistical features from a session grows with the input data length. In our tasks, deep NN algorithms become very effective as monitoring intervals get longer than $20$ seconds. 


\begin{figure*}[t]
	\centering
	\begin{subfigure}[t]{.8\columnwidth}
		\centering
		\includegraphics[width=0.98\columnwidth]{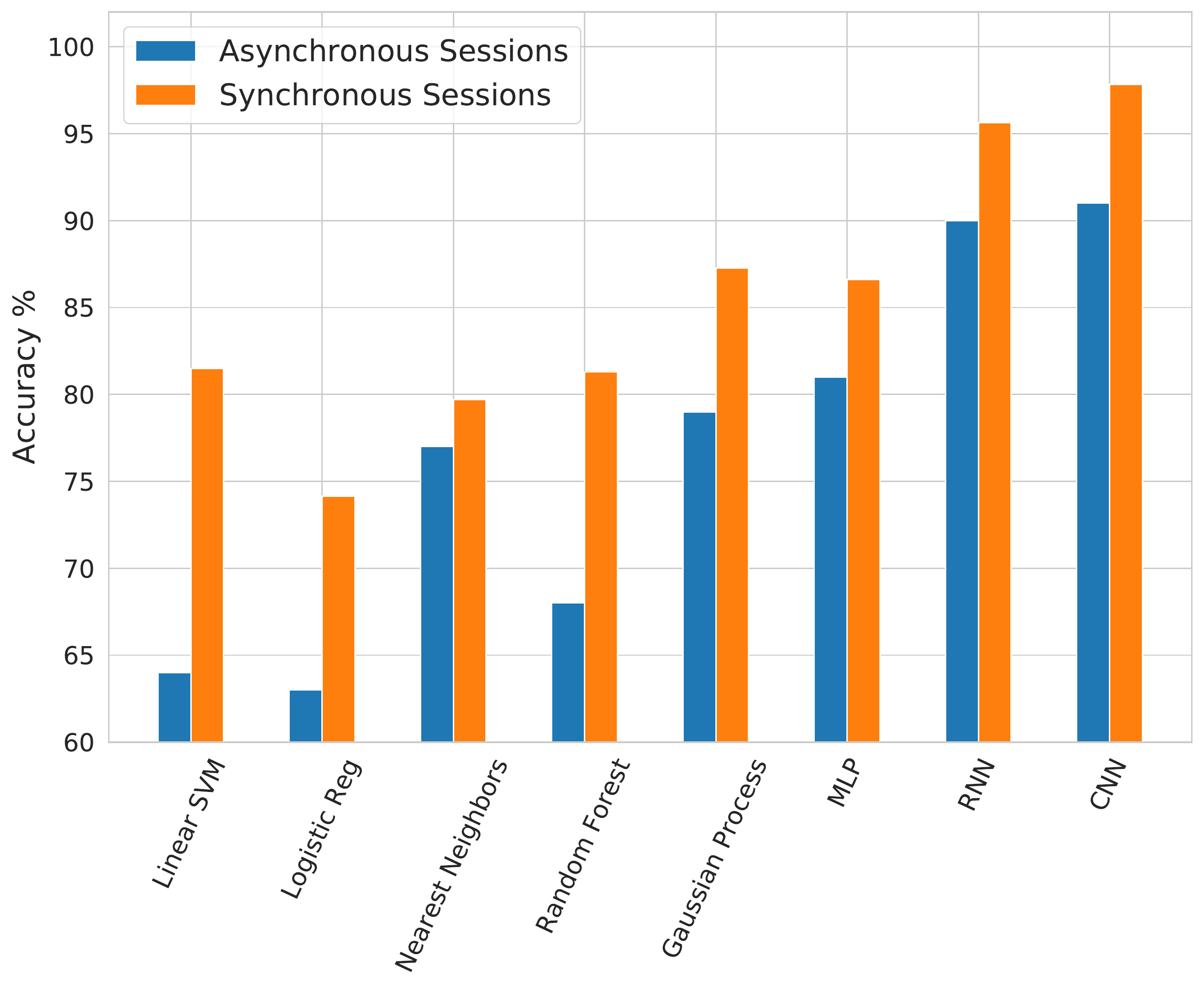}
		\caption{App identification.}
		\label{fig:sync_async_app}	
	\end{subfigure}
	\begin{subfigure}[t]{.8\columnwidth}
		\centering
		\includegraphics[width=0.98\columnwidth]{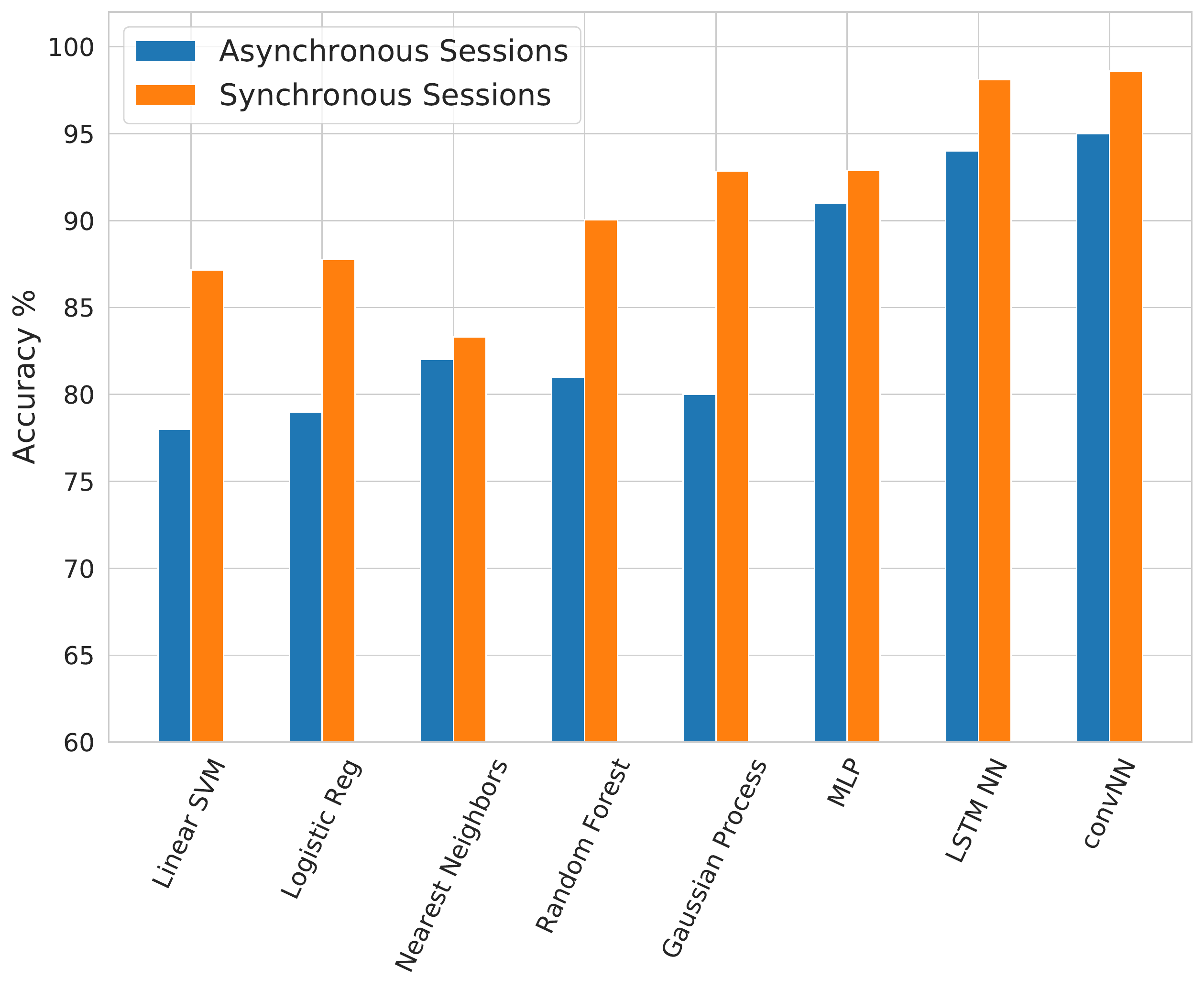}
		\caption{Service identification.}
		\label{fig:sync_async_app_service}	
	\end{subfigure}
	\centering
	\caption{
		Effect of Synchronization vs Asynchronization procedures}
	\label{fig:sync_async_app}
\end{figure*} 

\section{Unsupervised Traffic Profiling}
\label{sec:traffic_profilig}

Next, we analyze the mobile traffic exchanged within the four selected cell sites (see Section~\ref{sec:unlabeled_dataset}). The traffic load is modeled in terms of aggregated traffic dynamics and type of service requests over the $24$ hours of a day. The identification of mobile traffic, for each of the considered services, is performed using the trained classifiers from Section~\ref{sec:sl} with the {\it unlabeled} dataset. Formally, for each \ac{eNodeB}, the corresponding unlabeled dataset is stored into the tensor $\bm{X}^\prime$, with size $M^\prime \times N \times D$, where $M^\prime$ corresponds to the number of monitored \acp{RNTI} (sessions,  $N$ to the number of collected samples per session and $D=2$ is the number of communications directions (downlink and uplink). Given $\bm{X}^\prime$, as input, the classifier $c$ computes the output $\bm{Y}^\prime$, whose analysis provides a detailed characterization of the mobile user requests for the \ac{eNodeB} within the monitored time span. Vectors $\bm{x}^\prime_m$ and $\bm{y}^\prime_m = c(\bm{x}^\prime_m)$ respectively indicate the \mbox{$m$-th} sample of $\bm{X}^\prime$ and $\bm{Y}^\prime$. In this paper, we restrict our attention to the unsupervised classification of services, and use the \ac{CNN} classifier, as it yields the highest accuracy.

\subsection{Aggregated Traffic Analysis}

\begin{figure}[h]
	\centering
	\includegraphics[width=.8\columnwidth]{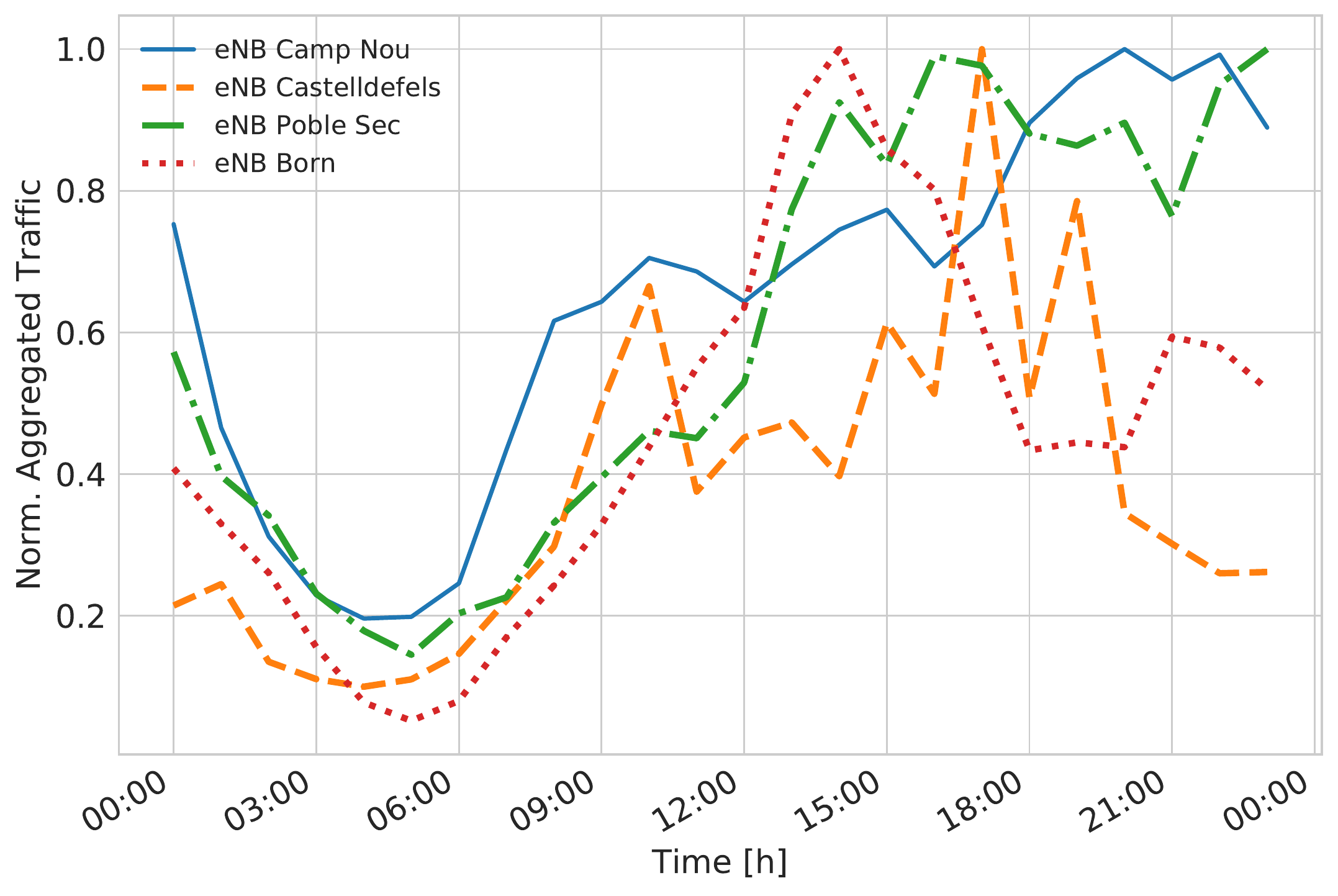}
	\caption{Daily Aggregated Traffic for the four eNodeBs.}
	\label{fig:aggregate_traffic}
\end{figure}

Fig.~\ref{fig:aggregate_traffic} shows the {\it aggregated} traffic demand for the four selected \acp{eNodeB} over the $24$ hours of a typical day, where each curve has been normalized with respect to the maximum traffic peak occurred during the day for the corresponding \ac{eNodeB}.
The four traffic profiles have a different trend, which depends on the characteristics of the served area (demographics, predominant land use, etc.), as confirmed by~\cite{furno2017tale}. \textbf{PobleSec} is a residential neighborhood and, as such, presents traffic peaks during the evening, at 5 and 11~pm. \textbf{Born} is instead a downtown district with a mixed residential, transport and leisure land use. Two peaks are detected: the highest is at lunchtime around 2~pm, whereas the second one is at dinner time, from 9~pm. This traffic behavior is likely due to the many restaurants and bars in the area. \textbf{CampNou} is mainly residential and presents a similar profile to PobleSec. However, Barcelona FC stadium is located in this area, and three football matches took place during the monitored period (events started at 8:45~pm and ended at 10:45~pm). As expected, a higher traffic intensity is observed during the football match hours. In particular, we registered a high amount of traffic exchanged between $7$~pm to $1$~am, i.e., before, during and after the events. This behavior is probably due to the movement of people attending the matches. \textbf{Castelldefels} is a suburban and low populated area with a university campus. The traffic variation suggests a typical office profile with traffic peaks at 10~am and 5~pm. However, in this radio cell the amount of traffic exchanged is the lowest observed across all \ac{eNodeB} sites, i.e., $6.8$~Gb/hour in the peak hours. The highest traffic intensity was measured in Camp Nou, reaching a peak of $106.1$~Gb/hour ($29.5$~Mb/s on average). Intermediate peak values are detected in Poble Sec and Born, amounting to $49.7$~Mb/s and $46.1$~Mb/s, respectively. The only common pattern among the four areas is the low traffic intensity at night, between $2$~am and $7$~am.

\subsection{Traffic Decomposition at Service Level}

The set of applications that we have labeled is restricted to those apps and services that dominate the radio resource usage. However, additional apps may also be present in the monitored traffic, such as Facebook, Instagram, Snapchat, etc. These apps generate mixed content, including \mbox{audio-streaming}, \mbox{video-streaming} and \mbox{video-calling}. Additional service types may also be generated by, e.g., \mbox{web-browsing} and file downloading. While in the present work the classifiers were not trained to specifically track these apps, for a robust classification outcome, it is desirable that the audio and video streams that they generate will either be captured and classified into the correct service class, or at least flagged as unknown. To locate those traffic patterns for which our classifier may produce inaccurate results, in our analysis we additionally account for the detection of \mbox{out-of-distribution (OOD)} sessions, i.e., of DCI traces that show different traffic dynamics from those learned at training time. To identify these ``statistical outliers'', the DCI data from each new session, $\bm{x}^\prime_m$, is fed to the \ac{CNN} and the corresponding softmax output vector \mbox{$\bm{y}^\prime_m = [y^\prime_{m1}, \dots, y^\prime_{mK}]^T$} is used to discriminate whether $\bm{x}^\prime_m$ is OOD or not, following the rationale in~\cite{hendrycks2018deep}\cite{sigurdsson2002outlier}. In detail, the \mbox{$k$-th} softmax output corresponds to the probability estimate that a given input session $\bm{x}^\prime_m$ belongs to class $k$, i.e., \mbox{$y^\prime_{mk} =\textrm{Prob}(\bm{x}^{\prime}_m \in \text{class \textit{k}})$}, with $k=1,\dots,K$. The classifier chooses the class $k^\star$ that maximizes this probability, i.e., 
\begin{equation}
\label{eq:kstar}
k^\star = \arg\!\max_k y^\prime_{mk}.
\end{equation} 
If a new app, not considered in the training phase, generates sessions having similar characteristics to those in the training set, namely, \mbox{audio-streaming}, \mbox{video-streaming} or real time \mbox{video-calls}, we expect the \ac{CNN} to generalize well and return similar vectors at the output of the softmax layer. That is, the softmax vector that is outputted at runtime for the new app should be sufficiently ``close'' to the output learned by the classifier from the labeled dataset, as the new signal bears statistical similarities with those learned in the training phase. In this case, it makes sense to accept the session and classify it as belonging to class $k^\star$. Otherwise, the session would be classified as OOD.

The problem at hand, boils down to assessing whether the softmax output $\bm{y}^\prime_m$ belongs to the statistical distribution learned by the \ac{CNN} or it is an outlier. This amounts to performing outlier detection in a multivariate setting, with $\bm{y}^\prime_m \in [0,1]^K$, $\sum_k y^\prime_{mk} = 1$. Among the many algorithms that can be used to this purpose, we adopt the method based on Kernelized Spatial Depth (KSD) functions of~\cite{Chen-2009} as it is lightweight and does not require the direct estimation of the probability density function (pdf) of the softmax output layer, which is a critical point, as good estimates require training over many points. Briefly, for a vector $\bm{y} \in \mathbb{R}^K$, we define the spatial sign function as $S(\bm{y}) = \bm{y}/\|\bm{y}\|$ if $\bm{y} \neq \bm{0}$ and $S(\bm{y}) = \bm{0}$ if $\bm{y}=\bm{0}$, where $\| \bm{y} \| = (\bm{y}^T \bm{y})^{1/2}$ is the \mbox{norm-2}. If $\mathcal{Y}_k$ is a training set containing $\ell$ softmax output vectors for a certain class $k$, $\mathcal{Y}_k = \{\bm{y}^{(k)}_1, \bm{y}^{(k)}_2, \dots, \bm{y}^{(k)}_\ell \}$, the {\it sample spatial depth} associated with a new softmax output vector $\bm{y}^\prime_m$ is:
\begin{equation}
\label{eq:spatial_depth}
D(\bm{y}^\prime_m,\mathcal{Y}_k) = 1 - \frac{1}{|\mathcal{Y}_k \cup \bm{y}^\prime_m|-1} \left \|\sum_{\bm{y} \in \mathcal{Y}_k} S(\bm{y}-\bm{y}^\prime_m) \right \| .
\end{equation}
Note that $D(\bm{y}^\prime_m,\mathcal{Y}_k) \in [0,1]$ provides a {\it measure of centrality} of the new point $\bm{y}^\prime_m$ with respect to the points in the training set $\mathcal{Y}_k$. In particular, if $D(\bm{y}^\prime_m,\mathcal{Y}_k)=1$, it follows that $\|\sum_{\bm{y} \in \mathcal{Y}_k} S(\bm{y}-\bm{y}^\prime_m) \| = 0$ and the new point is said to be the {\it spatial median} of set $\mathcal{Y}_k$, i.e., it can be thought of as the ``center of mass'' of this set. Hence, the spatial depth attains the highest value of $1$ at the spatial median and decreases to zero as $\bm{y}^\prime_m$ moves away from it. The spatial depth can thus be used as a measure of ``extremeness'' of a new data point with respect to a set. In~\cite{Chen-2009}, the spatial depth of Eq.~(\ref{eq:spatial_depth}) is {\it kernelized}, which means that distances are evaluated using a positive definite kernel map. A common choice, that we also use in this paper, is the generalized Gaussian kernel $\kappa(\bm{x},\bm{y})$,
\begin{equation}
\label{eq:kernel}
\kappa(\bm{x}, \bm{y}) = \exp(-(\bm{x} - \bm{y})^T \bm{\Sigma}^{-1} (\bm{x} - \bm{y})),
\end{equation}
which provides a measure of similarity between $\bm{x}$ and $\bm{y}$. Noting that the square norm can be expressed as
\begin{equation}
\label{eq:norm}
 \left \|\bm{x}-\bm{y}\right \|^2 = \bm{x}^T\bm{x}+\bm{x}^T\bm{x}-2\bm{x}^T\bm{y},
\end{equation}
kernelizing the sample spatial depth amounts to expanding (\ref{eq:spatial_depth}) and replacing the inner products with the kernel function $\kappa$. This returns the \textit{sample KSD function} (Eq.~(4) in~\cite{Chen-2009}).\\

\noindent \textbf{Session classification procedure in an unsupervised setting:} the \ac{CNN} classifier is augmented through the detection of OOD sessions, as follows:
\begin{itemize}
\item \textit{Initialization:} for each class $k=1,\dots,K$ in the service/app identification task a number of softmax output vectors is computed by the {\it trained} \ac{CNN} using the sessions in the training set. These softmax vectors are stored in the set $\mathcal{Y}_k$. Note that, being the results of a supervised training of the \ac{CNN}, we know that the vectors in $\mathcal{Y}_k$ are all generated by a distribution that is correctly learned during the supervised training phase. 
\item \textit{Feature extraction through the \mbox{pre-trained} \ac{CNN}:} at runtime, as a new \ac{DCI} vector $\bm{x}^\prime_m$ is measured, it is inputted into the \mbox{pre-trained} \ac{CNN}, obtaining the corresponding softmax output $\bm{y}^\prime_m$.
\item \textit{Classification and OOD detection:} vector $\bm{y}^\prime_m$ is used with Eq.~(\ref{eq:kstar}) to assess the most probable class $k^\star$. At this point, Algorithm~1 of~\cite{Chen-2009} is utilized to assess whether $\bm{y}^\prime_m$ is an outlier. In case the vector is classified as an outlier, it is assigned to the OOD class, otherwise it is assigned to class $k^\star$.    
\end{itemize}
Some final remarks are in order. The outlier detection algorithm uses a threshold $t \in [0,1]$, which allows exploring the tradeoff between {\it false alarm} rate and {\it detection} rate. Instead, the covariance matrix $\bm{\Sigma}$ controls the decision boundary for rejecting vectors, driving the tradeoff between the local and global behavior of KSD. If properly chosen, the contours of KSD should closely follow those of the (actual) underlying statistical distribution. $\bm{\Sigma}$ is learned, for each class $k$, from the training vectors in $\mathcal{Y}_k$, and for the following results we picked $\bm{\Sigma} = \bm{\Sigma}_2$ in~\cite{Chen-2009}.\\

\begin{figure}[t]
	\centering
	\includegraphics[width=.8\columnwidth]{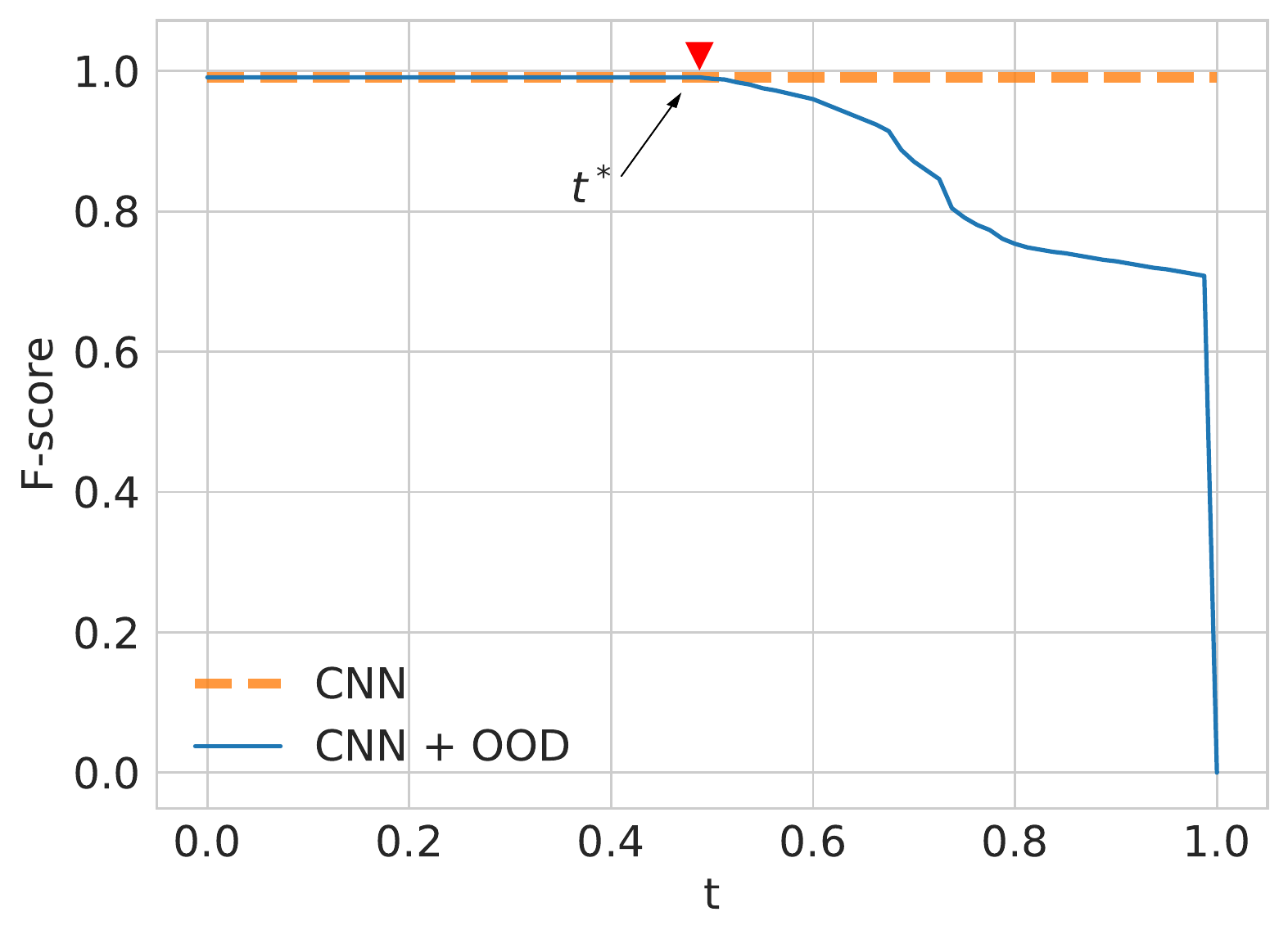}
	\caption{Finding threshold $t^\star$ using the CNN with (solid line) and without (dashed line) the OOD detection mechanism.}
	\label{fig:optimal_t}
\end{figure}

\noindent \textbf{Tuning the OOD threshold $t$:} for each class $k$, $\mathcal{S}_k$ is obtained from the training dataset. We recall that $\mathcal{S}_k$ is used to compute the covariance matrix associated with the adopted Gaussian kernel, which models the contours of the pdf of the output softmax vectors. The threshold $t \in [0,1]$ is instead used by the outlier detection algorithm to gauge the (kernelized) distance between the center of mass of set $\mathcal{S}_k$ and a new softmax vector, acquired at runtime. If $t=1$, the kernelized spatial depth of the new point will always be smaller than or equal to $t$ and all points will be rejected (marked as outliers). This is of course of no use. However, as we decrease $t$ towards $0$, we see that more and more points will be accepted, until, for $t=0$, no rejections will occur. So, $t$ determines the selectivity of the outlier detection mechanism, the higher $t$, the more selective the algorithm is. For our numerical evaluation, once the sets $\mathcal{S}_k$ are obtained for all classes $k$, we set this threshold by picking the highest value of it, $t^\star$, for which all the softmax vectors belonging to the test set are accepted, i.e., none of them is marked as an outlier (OOD). In other words, this is equivalent to making sure that the $F$-Score obtained over the test set from our trained \ac{CNN} without the OOD mechanism enabled equals that of the \ac{CNN} classifier augmented with the OOD detection capability. As $t^\star$ is the highest value of $t$ for which all the data in the test set are correctly classified as valid, our approach amounts to tuning the threshold in such a way that the outlier detection algorithm will be as selective as possible, while correctly treating all the data in the test set. In Fig.~\ref{fig:optimal_t}, we show the \mbox{$F$-Score} as a function of $t$ for the \ac{CNN} algorithm with and without OOD detection. Threshold $t^\star=0.48$ corresponds to the highest value of $t$ for which the \mbox{$F$-Score} remains at its maximum, i.e., at the end of the flat region.\\

\begin{figure*}[h]
	\centering
	\includegraphics[width=.9\columnwidth]
	{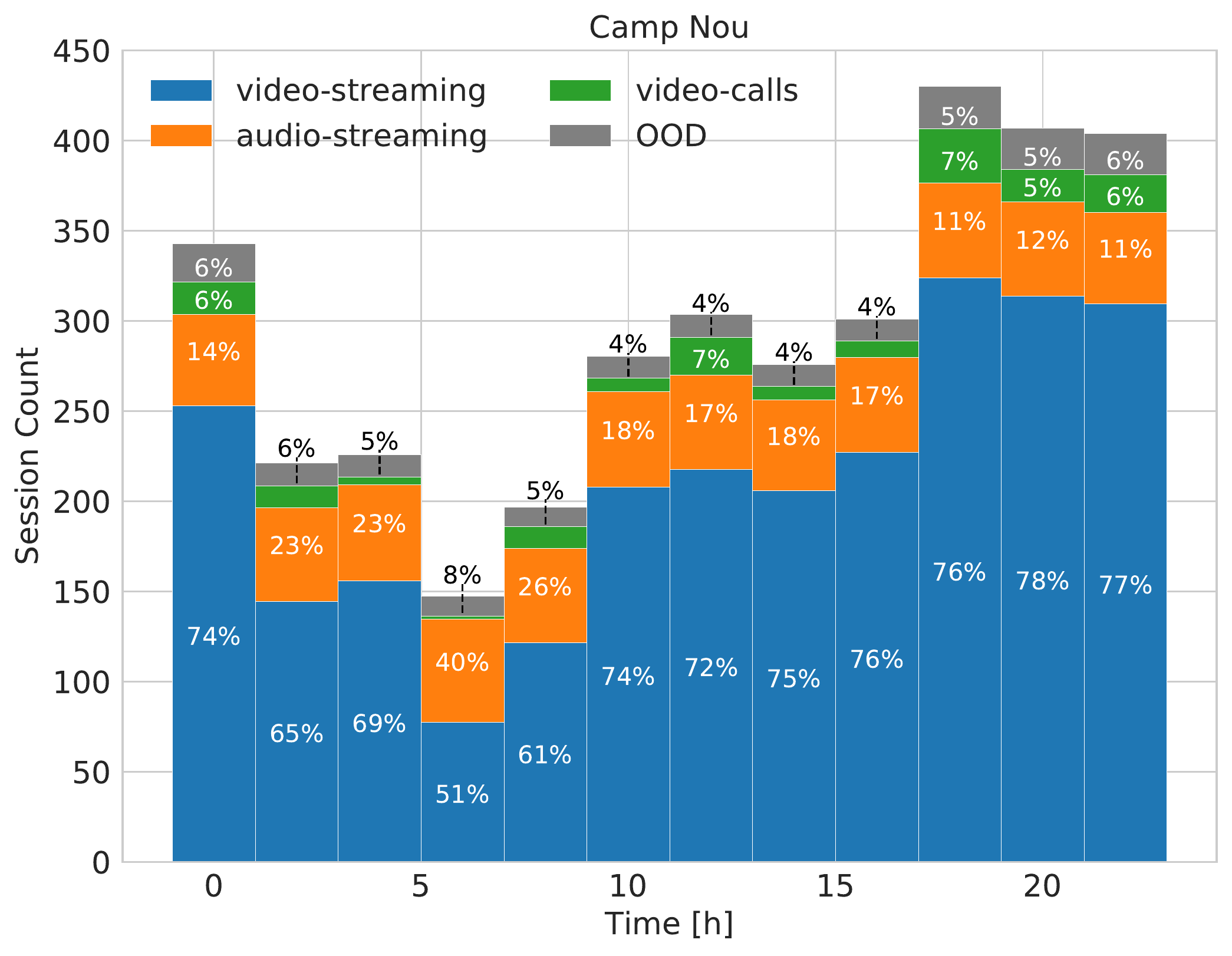}
	\includegraphics[width=.9\columnwidth]
	{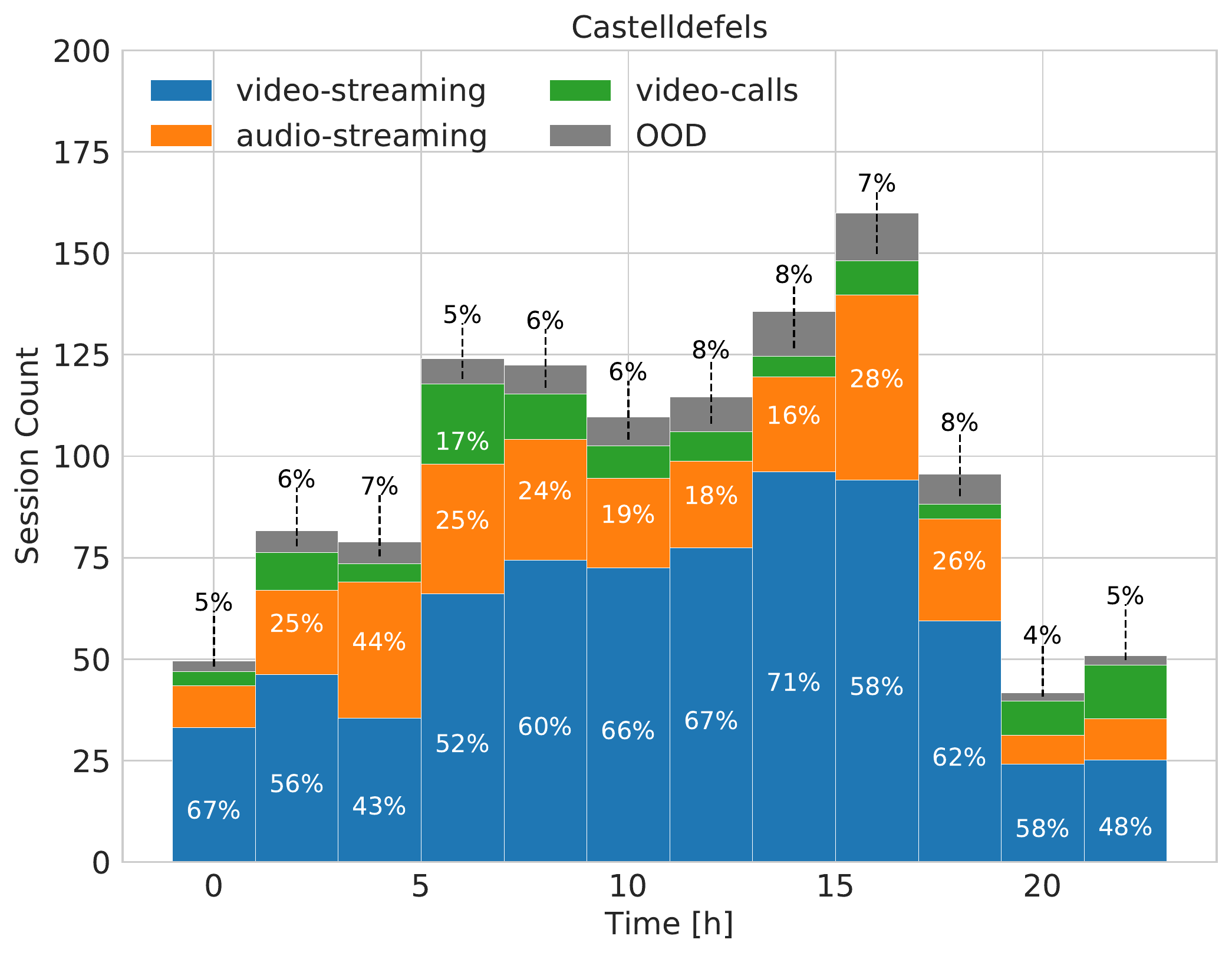}
	\includegraphics[width=.9\columnwidth]
	{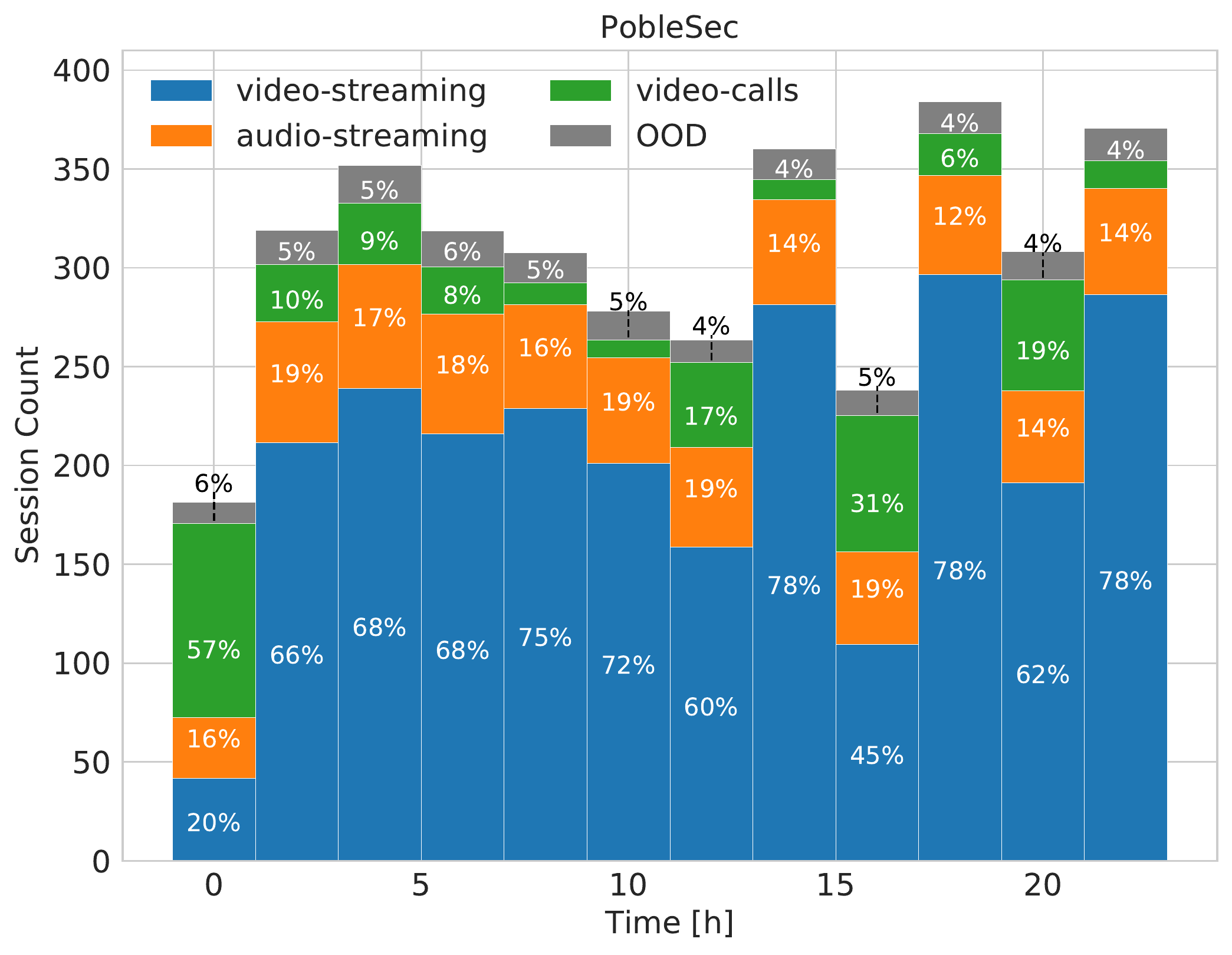}
	\includegraphics[width=.9\columnwidth]
	{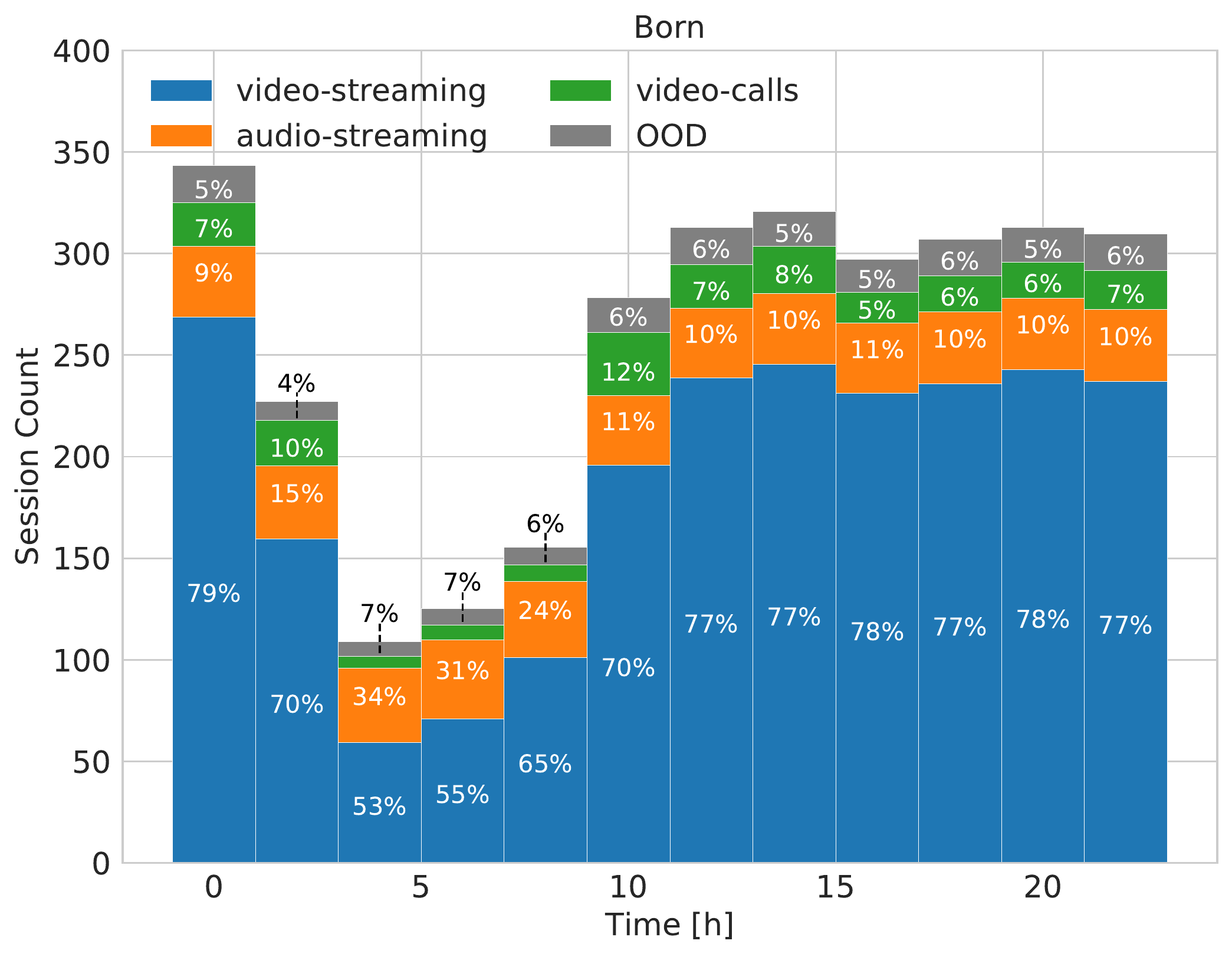}
	\caption{Traffic decomposition at service level for the four monitored \acp{eNodeB} during the $24$ hours of a day.}
	\label{fig:traffic_composition}
\end{figure*}

\noindent \textbf{Experimental analysis of \ac{eNodeB} traffic:} in Fig.~\ref{fig:traffic_composition}, the traffic decomposition into the considered service classes is shown for the four selected \acp{eNodeB} using $t^\star=0.48$. The percentage of sessions identified as OOD, for which the classifier is uncertain, is also reported at the top of each bar.
Common characteristics are observed in all the considered deployments:
\begin{itemize}
	\item the most used service is \mbox{video-streaming}, with typical shares ranging from $50\%$ to $80\%$. This confirms the measurements in~\cite{ericsson} and~\cite{ciscovni}.
	\item The least used service is \mbox{video-call}, whose share is typically between $5\%$ and $10\%$, whereas \mbox{audio-streaming} takes $21\%$ of the total traffic load.
	\item OOD sessions are consistently well below $8\%$. Note that this share accounts for all those apps that are not tracked by our classifier, such as texting, web browsing, and file transfers. 
\end{itemize}

Through the proposed service identification approach, we can accurately characterize, at runtime, the used services. Moreover, the traffic decomposition at service level allows one to make some interesting considerations on the land use. For example, in a typical residential area (PobleSec) the \mbox{audio-streaming} service is the one used the least across the four monitored sites, with an average of $16.4\%$. Instead, in a typical office and university neighborhood (Castelldefels), \mbox{audio-streaming} has the highest traffic share across all sites ($22\%$ on average). Born and CampNou, which are two leisure districts, present a similar traffic distribution across the day.  We finally remark that, while the traffic profiling results are shown using a time granularity of one hour, our classification tool allows for traffic decomposition at much shorter timescales, i.e., on a \mbox{per-session} basis.

\section{Related Work}
\label{sec:related}

The most common classification methods in the literature leverage UDP/TCP port analysis and/or packet inspection. 

UDP/TCP port analysis relies on the fact that most Internet applications use \mbox{well-known} \ac{TCP} or \ac{UDP} port numbers. For instance, the authors of~\cite{choi2012automated} define a mobile traffic classifier as a collection of rules, including destination \ac{IP} addresses and port numbers. Based on these rules, \mbox{application-level} mobile traffic identification is performed deploying a dedicated classification architecture within the network, and measurement agents at the mobile devices. However, \mbox{port-based} schemes hardly work in the presence of applications using dynamic port numbers~\cite{fu2016service}. 

A scheme based on deep packet inspection is presented in~\cite{yang2012empirical}. The authors of this paper devise a technique for \ac{CDMA} traffic classification, using \mbox{correlation-based} feature selection along with a decision tree classifier trained on a  labeled dataset (which is labeled via deep packet inspection). The algorithm in~\cite{han2012maximum} extracts application layer payload patterns, and performs maximum \mbox{entropy-based} \mbox{IP-traffic} classification exploiting different \ac{ML} algorithms such as Naive Bayes, \acp{SVM} and partial decision trees. Remarkably, \mbox{payload-based} methods are limited by a significant complexity and computation load~\cite{fu2016service}. Furthermore, many mobile applications adopt encrypted data transmission due to security and privacy concerns, which renders packet inspection approaches ineffective. 

Classification schemes for encrypted flows utilize traffic statistics, extracting meaningful features from the observed traffic patterns. For example, the authors of~\cite{liu2018cascade} propose a classification system for mobile apps based on a Cascade Forest algorithm exploiting features related to: i) the mobile traces, ii) the Cascade Forest algorithm, iii) \mbox{connection-oriented} protocols and \mbox{connection-less} protocols, and iv) encrypted and \mbox{unencrypted} flows. Along the same lines, a classification approach that combines \mbox{state-of-art} classifiers for encrypted traffic analysis is put forward in~\cite{aceto2018mobile}.

Another interesting work is presented in~\cite{fu2016service}, where the authors classify service usage from mobile messaging apps by jointly modeling user behavioral patterns, network traffic characteristics, and temporal dependencies. The framework is built upon four main blocks: traffic segmentation, traffic feature extraction, service usage prediction, and outlier detection. When traffic flows are short and the defined features do not suffice to fully describe the traffic pattern, \acp{HMM} are exploited to capture temporal dependencies, to enhance the classification accuracy.

The authors of~\cite{2016eavesdropping} show that a passive eavesdropper is capable of identifying fine grained user activities for \mbox{Android} and iOS mobile apps, by solely inspecting \ac{IP} headers. Their technique is based on the intuition that the highly specific implementation of an app may leave a fingerprint on the generated traffic in terms of, e.g., transfer rates, packet exchanges, and data movement. For each activity type, a behavioral model is built, then \mbox{K-means} and \ac{SVM} are respectively used to reveal which model is the most appropriate, and to classify the mobile apps. 

Automatic fingerprinting and \mbox{real-time} identification of Android apps from their encrypted network traffic is presented in~\cite{taylor2016appscanner}. \mbox{IP-based} feature extraction and supervised learning algorithms are the basis of a framework featuring six classifiers, obtained as variations of \acp{SVM} and \acp{RF}. \acp{RF} have also been considered in~\cite{2015enhancing}, where the authors claim that the sole use of \mbox{packet-based} features does not suffice to classify the traffic generated by mobile apps. As a solution, they  use a combination of packet size distributions and communication patterns. 

Recent works exploit \acp{NN}~\cite{chen2016automatic}\cite{zhang2017zipnet}\cite{aceto2018mobile}. In~\cite{chen2016automatic}, mobile apps are identified by automatically extracting features from labeled packets through \acp{CNN}, which are trained using raw \ac{HTTP} requests. In~\cite{aceto2018mobile}, encrypted traffic is classified using deep learning architectures (feed forward, convolutional and recurrent neural networks) for \mbox{Android} and iOS mobile apps, with and without using TCP/UDP ports. The authors of~\cite{zhang2017zipnet} combine Zipper Networks (ZipNet) and \ac{GAN} to infer narrowly localized and fine grained traffic generation from coarse measurements.

A systematic framework is devised in~\cite{aceto2019mobile} for comparison among different techniques where deep learning is proposed as the most viable strategy. The performance of the deep learning classifiers is thoroughly investigated based on three mobile datasets of real human users’ activity, highlighting the related drawbacks, design guidelines, and challenges.
Several survey papers dealing with deep learning techniques applied to traffic classification can be found in~\cite{rezaei2019deep}, ~\cite{wang2019survey} and~\cite{zhang2019deep}. The authors in~\cite{rezaei2019deep} overview general guidelines for classification tasks, present some deep learning techniques and how they have been applied for traffic classification, and finally, open problems and future directions are addressed. 
The survey in~\cite{wang2019survey} presents a deep learning-based framework for mobile encrypted traffic classification, reviewing existing work according to dataset selection, model input design, and model architecture, and highlighting open issues and challenges.
Finally, a comprehensive and thorough study of the crossovers between deep learning and mobile networking research is provided in~\cite{zhang2019deep} where the authors discuss how to tailor deep learning to mobile environments. Current challenges and open future research directions are also discussed. 

We stress that that most of the works in the literature, with the exception of~\cite{chen2016automatic,zhang2017zipnet,aceto2018mobile} and~\cite{aceto2019mobile}, classify mobile traffic based on manual feature extraction and all the papers that we surveyed process network or application level data. Our work departs from prior art as we leverage the feature extraction capabilities of deep neural networks and classify mobile data gathered from the physical channel of a mobile operative network, at runtime, and without access to application data and TCP/UDP port numbers.

\section{Conclusions}
\label{sec:conclusions}

In this paper, we have presented a framework that allows highly accurate classification of application and services from \mbox{radio-link} level data, at runtime, and without having to decrypt dedicated physical layer channels. To this end, we decoded the LTE Physical Downlink Control Channel (\ac{PDCCH}), where Downlink Control Information (\ac{DCI}) messages are sent in clear text. Through DCI data, it is possible to track the data flows exchanged between the serving cell and its active users, extracting features that allow the reliable identification of the apps/services that are being executed at the mobile terminals. For the classification of such traffic, we have tailored deep artificial Neural Networks \acp{NN}, namely, Multi-Layer Perceptron (MLP), Recurrent NNs and Convolutional NNs, comparing their performance against that of benchmark classifiers based on \mbox{state-of-the-art} supervised learning algorithms. Our numerical results show that NN architectures overcome the other approaches in terms of classification accuracy, with the best accuracy (as high as $98$\%) being achieved by \acp{CNN}. As a major contribution of this work, labeled and unlabeled datasets of DCI data from real radio cell deployments have been collected. The labeled dataset has been used to train and compare the classifiers. For the unlabeled dataset, we have augmented the \ac{CNN} with the capability of detecting input DCI data that do not conform to that learned during the training phase: the corresponding patterns are detected and associated with an {\it unknown} class. This increases the robustness of the \ac{CNN} classifier, allowing its use, at runtime, to perform fine grained traffic analysis from radio cell sites from an operative mobile network. To summarize, the main outcomes of our work are: \textbf{1)} a methodology to extract DCI data from the PDCCH channel, and for the use of such data to train traffic classifiers, \textbf{2)} the fine tuning and a thorough performance comparison of classification algorithms, \textbf{3)} the design of a novel technique for the {\it fine grained} and {\it online} traffic analysis of communication sessions from real radio cell sites, and \textbf{4)} the discussion of the traffic distribution resulting from such analysis from four selected sites of a Spanish mobile operator, in the city of Barcelona. 


\bibliographystyle{IEEEtran}
\bibliography{biblio}

\begin{thebibliography}{10}
\providecommand{\url}[1]{#1}
\csname url@samestyle\endcsname
\providecommand{\newblock}{\relax}
\providecommand{\bibinfo}[2]{#2}
\providecommand{\BIBentrySTDinterwordspacing}{\spaceskip=0pt\relax}
\providecommand{\BIBentryALTinterwordstretchfactor}{4}
\providecommand{\BIBentryALTinterwordspacing}{\spaceskip=\fontdimen2\font plus
\BIBentryALTinterwordstretchfactor\fontdimen3\font minus
  \fontdimen4\font\relax}
\providecommand{\BIBforeignlanguage}[2]{{%
\expandafter\ifx\csname l@#1\endcsname\relax
\typeout{** WARNING: IEEEtran.bst: No hyphenation pattern has been}%
\typeout{** loaded for the language `#1'. Using the pattern for}%
\typeout{** the default language instead.}%
\else
\language=\csname l@#1\endcsname
\fi
#2}}
\providecommand{\BIBdecl}{\relax}
\BIBdecl

\bibitem{ericsson}
\BIBentryALTinterwordspacing
Ericsson. (2018) Ericsson mobility report june 2018. [Online]. Available:
  \url{https://www.ericsson.com/en/mobility-report/reports/june-2018}
\BIBentrySTDinterwordspacing

\bibitem{ciscovni}
\BIBentryALTinterwordspacing
Cisco. (2017) Cisco visual networking index: Global mobile data traffic
  forecast update, 2016–2021 white paper. [Online]. Available:
  \url{www.cisco.com/c/en/us/solutions/collateral/service-provider/visual-networking-index-vni/mobile-white-paper-c11-520862.html}
\BIBentrySTDinterwordspacing

\bibitem{chen2014requirements}
S.~Chen and J.~Zhao, ``The requirements, challenges, and technologies for 5g of
  terrestrial mobile telecommunication,'' \emph{{IEEE communications
  magazine}}, vol.~52, no.~5, pp. 36--43, 2014.

\bibitem{Laurila2012}
J.~K. Laurila, D.~Gatica-Perez, J.~B. I.~Aad, O.~Bornet, T.-M.-T. Do,
  O.~Dousse, J.~Eberle, and M.~Miettinen, ``The mobile data challenge: Big data
  for mobile computing research,'' in \emph{{Mobile Data Challenge Workshop
  (MDC), in conjunction with ``Pervasive 2012''}}, {Newcastle, UK}, 2012.

\bibitem{Barlacchi2015}
G.~Barlacchi, M.~D. Nadai, R.~Larcher, A.~Casella, C.~Chitic, G.~Torrisi,
  F.~Antonelli, A.~Vespignani, A.~Pentland, and B.~Lepri, ``{A multi-source
  dataset of urban life in the city of Milan and the Province of Trentino},''
  \emph{{Scientific Data}}, vol.~2, no. 150055, pp. 1--15, 2015.

\bibitem{Reades2009}
J.~Reades, F.~Calabrese, and C.~Ratti, ``Eigenplaces: analysing cities using
  the space-time structure of the mobile phone network,'' \emph{{Environment
  and Planning B: Planning and Design}}, vol.~36, pp. 824--836, 2009.

\bibitem{aceto2018mobile}
G.~Aceto, D.~Ciuonzo, A.~Montieri, and A.~Pescap{\'e}, ``Mobile encrypted
  traffic classification using deep learning,'' in \emph{{Network Traffic
  Measurement and Analysis Conference (TMA)}}.\hskip 1em plus 0.5em minus
  0.4em\relax {Vienna, Austria}: IEEE, June 2018.

\bibitem{zhang2017zipnet}
C.~Zhang, X.~Ouyang, and P.~Patras, ``Zipnet-gan: Inferring fine-grained mobile
  traffic patterns via a generative adversarial neural network,'' in
  \emph{{International Conference on emerging Networking EXperiments and
  Technologies (CoNEXT)}}.\hskip 1em plus 0.5em minus 0.4em\relax
  {Seoul-Incheon, South Korea}: ACM, 2017.

\bibitem{chen2016automatic}
Z.~Chen, B.~Yu, Y.~Zhang, J.~Zhang, and J.~Xu, ``Automatic mobile application
  traffic identification by convolutional neural networks,'' in \emph{{IEEE
  Trustcom/BigDataSE/ISPA}}.\hskip 1em plus 0.5em minus 0.4em\relax {Tianjin,
  China}: IEEE, August 2016.

\bibitem{bui2016owl}
N.~Bui and J.~Widmer, ``{Owl: A reliable online watcher for lte control channel
  measurements},'' in \emph{{Workshop on All Things Cellular: Operations,
  Applications and Challenges}}.\hskip 1em plus 0.5em minus 0.4em\relax {New
  York, NY, USA}: ACM, October 2016.

\bibitem{earth-D23}
{EU EARTH: Energy Aware Radio and neTwork tecHnologies}, ``{D2.3: Energy
  efficiency analysis of the reference systems, areas of improvements and
  target breakdown},'' {Deliverable D2.3, \url{www.ict-earth.eu}}, 2010.

\bibitem{Xu2017understanding}
F.~Xu, Y.~Li, H.~Wang, P.~Zhang, and D.~Jin, ``{Understanding Mobile Traffic
  Patterns of Large Scale Cellular Towers in Urban Environment},''
  \emph{{IEEE/ACM Transactions on Networking}}, vol.~25, no.~2, 2017.

\bibitem{gomez2016srslte}
I.~Gomez-Miguelez, A.~Garcia-Saavedra, P.~D. Sutton, P.~Serrano, C.~Cano, and
  D.~J. Leith, ``{srsLTE: an open-source platform for LTE evolution and
  experimentation},'' in \emph{{ACM International Workshop on Wireless Network
  Testbeds, Experimental Evaluation, and Characterization (WiNTECH)}}.\hskip
  1em plus 0.5em minus 0.4em\relax {New York, NY, USA}: ACM, October 2016.

\bibitem{etsi}
``E-{U}{T}{R}{A}; physical layer procedures,'' \emph{3GPP TS, vol. 36.213},
  2016.

\bibitem{bishop2006}
C.~M. Bishop, \emph{Pattern recognition and machine learning}.\hskip 1em plus
  0.5em minus 0.4em\relax Springer, 2006.

\bibitem{maas2013rectifier}
A.~L. Maas, A.~Y. Hannun, and A.~Y. Ng, ``Rectifier nonlinearities improve
  neural network acoustic models,'' in \emph{{International Conference on
  Machine Learning (ICML)}}, {Atlanta, USA}, June 2013.

\bibitem{tieleman2012lecture}
T.~Tieleman and G.~Hinton, ``{Divide the gradient by a running average of its
  recent magnitude},'' \emph{Neural networks for machine learning}, vol.~4,
  no.~2, pp. 26--31, 2012.

\bibitem{Hochreiter1997}
S.~Hochreiter and J.~Schmidhuber, ``{Long Short-Term Memory},'' \emph{{Neural
  Computation}}, vol.~9, no.~8, pp. 1735--1780, 1997.

\bibitem{gers1999learning}
F.~A. Gers, J.~Schmidhuber, and F.~Cummins, ``{Learning to forget: Continual
  prediction with LSTM},'' in \emph{{International Conference on Artificial
  Neural Networks (ICANN)}}.\hskip 1em plus 0.5em minus 0.4em\relax {Edinburgh,
  UK}: IEEE, 1999.

\bibitem{goodfellow2016deep}
I.~Goodfellow, Y.~Bengio, A.~Courville, and Y.~Bengio, \emph{{Deep
  Learning}}.\hskip 1em plus 0.5em minus 0.4em\relax {The MIT Press}, 2016.

\bibitem{gadaleta2016idnet}
M.~Gadaleta and M.~Rossi, ``{IDNet: Smartphone-based gait recognition with
  convolutional neural networks},'' \emph{{Pattern Recognition}}, vol.~74, pp.
  25--37, 2018.

\bibitem{fan2008liblinear}
R.-E. Fan, K.-W. Chang, C.-J. Hsieh, X.-R. Wang, and C.-J. Lin, ``Liblinear: A
  library for large linear classification,'' \emph{{Journal of machine learning
  research}}, vol.~9, no. Aug, pp. 1871--1874, 2008.

\bibitem{altman1992introduction}
N.~S. Altman, ``An introduction to kernel and nearest-neighbor nonparametric
  regression,'' \emph{The American Statistician}, vol.~46, no.~3, pp. 175--185,
  1992.

\bibitem{wu2007robust}
Y.~Wu and Y.~Liu, ``Robust truncated hinge loss support vector machines,''
  \emph{Journal of the American Statistical Association}, vol. 102, no. 479,
  pp. 974--983, 2007.

\bibitem{breiman2001random}
L.~Breiman, ``Random forests,'' \emph{Machine learning}, vol.~45, no.~1, pp.
  5--32, 2001.

\bibitem{rasmussen2004gaussian}
C.~E. Rasmussen, ``Gaussian processes for machine learning,'' in \emph{Advanced
  lectures on machine learning}.\hskip 1em plus 0.5em minus 0.4em\relax
  Springer, 2004, pp. 63--71.

\bibitem{furno2017tale}
A.~Furno, M.~Fiore, R.~Stanica, C.~Ziemlicki, and Z.~Smoreda, ``A tale of ten
  cities: Characterizing signatures of mobile traffic in urban areas,''
  \emph{{IEEE Transactions on Mobile Computing}}, vol.~16, no.~10, pp.
  2682--2696, 2017.

\bibitem{hendrycks2018deep}
D.~Hendrycks, M.~Mazeika, and T.~G. Dietterich, ``Deep anomaly detection with
  outlier exposure,'' in \emph{{International Conference on Learning
  Representations (ICLR)}}, {Vancouver, BC, Canada}, April 2018.

\bibitem{sigurdsson2002outlier}
S.~Sigurdsson, J.~Larsen, L.~K. Hansen, P.~A. Philipsen, and H.-C. Wulf,
  ``Outlier estimation and detection application to skin lesion
  classification,'' in \emph{{IEEE International Conference on Acoustics Speech
  and Signal Processing (ICASSP)}}.\hskip 1em plus 0.5em minus 0.4em\relax
  {Orlando, FL, USA}: IEEE, May 2002.

\bibitem{Chen-2009}
Y.~Chen, X.~Dang, H.~Peng, and H.~L. {Bart Jr.}, ``{Outlier Detection with the
  Kernelized Spatial Depth Function},'' \emph{{IEEE Transactions on Pattern
  Analysis and Machine Intelligence}}, vol.~31, no.~2, pp. 288--305, 2009.

\bibitem{choi2012automated}
Y.~Choi, J.~Y. Chung, B.~Park, and J.~W.-K. Hong, ``Automated classifier
  generation for application-level mobile traffic identification,'' in
  \emph{{IEEE Network Operations and Management Symposium (NOMS)}}.\hskip 1em
  plus 0.5em minus 0.4em\relax {Hawaii, USA}: IEEE, April 2012.

\bibitem{fu2016service}
Y.~Fu, H.~Xiong, X.~Lu, J.~Yang, and C.~Chen, ``{Service usage classification
  with encrypted internet traffic in mobile messaging apps},'' \emph{IEEE
  Transactions on Mobile Computing}, vol.~15, no.~11, pp. 2851--2864, 2016.

\bibitem{yang2012empirical}
J.~Yang, Z.~Ma, C.~Dong, and G.~Cheng, ``{An empirical investigation into CDMA
  network traffic classification based on feature selection},'' in
  \emph{{International Symposium on Wireless Personal Multimedia Communications
  (WPMC)}}.\hskip 1em plus 0.5em minus 0.4em\relax {Taipei, Taiwan}: IEEE,
  December 2012.

\bibitem{han2012maximum}
X.~Han, Y.~Zhou, L.~Huang, L.~Han, J.~Hu, and J.~Shi, ``{Maximum entropy based
  IP-traffic classification in mobile communication networks},'' in \emph{{IEEE
  Wireless Communications and Networking Conference (WCNC)}}.\hskip 1em plus
  0.5em minus 0.4em\relax {Shanghai, China}: IEEE, April 2012.

\bibitem{liu2018cascade}
Y.~Liu, S.~Zhang, B.~Ding, X.~Li, and Y.~Wang, ``A cascade forest approach to
  application classification of mobile traces,'' in \emph{{IEEE Wireless
  Communications and Networking Conference (WCNC)}}.\hskip 1em plus 0.5em minus
  0.4em\relax {Barcelona, Spain}: IEEE, April 2018.

\bibitem{2016eavesdropping}
B.~Saltaformaggio, H.~Choi, K.~Johnson, Y.~Kwon, Q.~Zhang, X.~Zhang, D.~Xu, and
  J.~Qian, ``{Eavesdropping on Fine-Grained User Activities Within Smartphone
  Apps Over Encrypted Network Traffic},'' in \emph{{USENIX Workshop on
  Offensive Technologies (WOOT)}}, {Austin, TX, USA}, August 2016.

\bibitem{taylor2016appscanner}
V.~F. Taylor, R.~Spolaor, M.~Conti, and I.~Martinovic, ``{Appscanner: Automatic
  fingerprinting of smartphone apps from encrypted network traffic},'' in
  \emph{{IEEE European Symposium on Security and Privacy}}.\hskip 1em plus
  0.5em minus 0.4em\relax {Saarbr\"{u}cken, Germany}: IEEE, March 2016.

\bibitem{2015enhancing}
S.~Mongkolluksamee, V.~Visoottiviseth, and K.~Fukuda, ``Enhancing the
  performance of mobile traffic identification with communication patterns,''
  in \emph{IEEE 39th Annual Computer Software and Applications Conference
  (COMPSAC)}.\hskip 1em plus 0.5em minus 0.4em\relax {Taichung, Taiwan}: IEEE,
  July 2015.

\bibitem{aceto2019mobile}
G.~Aceto, D.~Ciuonzo, A.~Montieri, and A.~Pescap{\'e}, ``Mobile encrypted
  traffic classification using deep learning: Experimental evaluation, lessons
  learned, and challenges,'' \emph{IEEE Transactions on Network and Service
  Management}, vol.~16, no.~2, pp. 445--458, 2019.

\bibitem{rezaei2019deep}
S.~Rezaei and X.~Liu, ``Deep learning for encrypted traffic classification: An
  overview,'' \emph{IEEE communications magazine}, vol.~57, no.~5, pp. 76--81,
  2019.

\bibitem{wang2019survey}
P.~Wang, X.~Chen, F.~Ye, and Z.~Sun, ``A survey of techniques for mobile
  service encrypted traffic classification using deep learning,'' \emph{IEEE
  Access}, vol.~7, pp. 54\,024--54\,033, 2019.

\bibitem{zhang2019deep}
C.~Zhang, P.~Patras, and H.~Haddadi, ``Deep learning in mobile and wireless
  networking: A survey,'' \emph{IEEE Communications Surveys \& Tutorials},
  vol.~21, no.~3, pp. 2224--2287, 2019.

\end{thebibliography}

\end{document}